\font\fta=cmbx10 scaled\magstep1
\font\ftb=cmmib10
\font\ftc=cmmib7
\font\ftd=cmbx10 scaled\magstep3
\font\fte=cmbsy10
\baselineskip=18pt
\centerline{\ftd Microscopic calculations of medium effects}
\centerline{\ftd for 200-MeV (p,p\raise.7ex\hbox{\fte 0}) reactions}
\vskip.4in
\centerline{F. Sammarruca}
\centerline{\it University of Idaho, Moscow, ID 83843 USA}
\vskip.4in
\centerline{E.J. Stephenson and K. Jiang}
\centerline{\it Indiana University Cyclotron Facility, Bloomington, IN
47408 USA}
\vfill
\centerline{\fta Abstract}
\vskip.4in
We examine the quality of a G-matrix calculation of the effective
nucleon-nucleon (NN) interaction for the prediction of the cross section
and analyzing power for 200-MeV (p,p$'$) reactions that populate natural
parity states in $^{16}$O, $^{28}$Si, and $^{40}$Ca.  This calculation is
based on a one-boson-exchange model of the free NN force that reproduces
NN observables well.  The G-matrix includes the effects of Pauli
blocking, nuclear binding, and strong relativistic mean-field potentials.
The implications of adjustments to the effective mass ansatz to improve
the quality of the approximation at momenta above the Fermi level will be
discussed, along with the general quality of agreement to a variety of
(p,p$'$) transitions.
\vfill
\eject
\noindent{\fta 1.~~Introduction}
\bigskip 
One of the goals of nuclear physics is to describe the structure
of nuclei and the dynamics of nuclear reactions in terms of the 
underlying interaction between nucleons. But from a practical 
standpoint a microscopic calculation of the many-body system is 
not a viable option. Thus a major quest in studies of the nuclear
many-body problem is the construction of an effective nucleon-nucleon  
interaction, modified so as to account for the presence of the 
many-body environment, that can be used in a simplified treatment of the 
properties of nuclear matter or nuclear reactions. 

The most commonly included many-body mechanisms arise from 
Pauli blocking and nuclear binding, and are the main aspects 
of what is known as the Brueckner G-matrix approach [1-6]. Its
relativistic extension in which the nuclear mean field is characterized
by strong, competing scalar and vector fields, is known as Dirac-Brueckner
theory [7-10] and has also become a rather established method.
This scheme accounts in a more natural way for the size of the spin-orbit
splitting seen in nuclear states. We will refer to all of these effects 
as ``conventional" medium modifications.

More exotic medium effects have been proposed. Perhaps the most extensively
discussed are those arising from changes in the QCD vacuum induced by increasing 
nuclear matter density, 
which are expected to lower both nucleon and 
meson effective masses in the medium [11,12]. Based in part on considerations
of quark condensates in nuclear matter [13], this reduction is expected to
be about 20\% for the meson masses at normal nuclear matter densities,
and would represent a precursor of the transition to a chirally restored
phase at higher temperatures and densities. Since the pion is a Goldstone boson, its mass would change only very slowly with increasing density.
For the tensor interaction 
which arises from the competing contributions of the 
long-range pion and the short-range
$\rho$-meson, a lowering of the $\rho$-meson mass would result
in a reduced tensor attraction at medium
range. This should be observable in nucleon-induced
reactions such as (p,p$'$) inelastic scattering or (p,n) charge exchange
through a measurement of the degree of spin-flip in polarization transfer
[14,15]. Charge-exchange quasi-free scattering consistently shows a lowering
of the ratio of spin-longitudinal to transverse responses relative to RPA
calculations, a result that has been interpreted as evidence for a weakening
of the pionic component in nuclei [16].  Comparable
results have been reported for discrete isovector transitions that also
suggest a weakened tensor force. The polarization transfer coefficient
$D_{NN}$, normally negative in the plane-wave limit [17], moves toward
zero [18,19] or becomes positive [20], for example.

In this paper, we will establish a basis from which the question of 
exotic contributions to the medium-dependent effective interaction
can be explored, in particular through the spin dependence of proton
elastic scattering, (p,p$'$) inelastic scattering, and (p,n) reactions.
We will begin with a one-boson-exchange (OBE) representation of the 
nucleon-nucleon (NN) interaction that provides an excellent reproduction
of the NN observables below pion production threshold. Then
the density-dependent
effective interaction is obtained by solving the Bethe-Goldstone 
equation in infinite nuclear matter.  It contains all the conventional 
effects as defined above.  Next
the resulting G-matrix is converted to a Yukawa-function form [21]
for use in Distorted Wave Impulse Approximation (DWIA) calculations.
The early sections of this paper describe these models, most of which 
have been explained in greater detail elsewhere, except for the transformation
from t-matrix to Yukawa-function form. This information is included here to 
support the observations and discussion later in the paper, some of which 
are unique to our application. Section 2 describes the OBE model, which is
an updated version of the Bonn interaction [22], and Section 3 outlines the 
G-matrix solution. The transformation is described in Section 4, and shown
in complete form in the Appendix. Details of the DWIA calculations are
given in Section 5. For the calculations in this paper we will make use 
of the LEA program from Kelly [23]. This program operates in coordinate
space and assumes zero-range exchange. The effect of this will be discussed
in more detail later in the paper.

The treatment of conventional medium effects through a G-matrix for use 
in (p,p$'$) or (p,n) reactions is certainly not new. Pauli blocking and
nuclear binding effects have been used to create interactions that have 
received wide examination [24-27]. While producing effects on the NN
interaction that are qualitatively similar, the size of the predicted 
changes differs from author to author. We will examine the 
central issues involved in the so-called ``effective mass approximation,"
a crucial step in the G-matrix method.
From studies of nuclear matter, effective
in-medium nucleon masses have been 
previously determined [22]. However, for nucleons with energies in the 
continuum, we find that this approximation works best for a different
choice of the effective masses. The consequences of this will be explored
in Section 6. There we will also illustrate how variations in this 
approximation 
affect the calculation of a folding model (t$\rho$) optical potential
and (p,p$'$) observables. 

Relatively few relativistic calculations of medium effects have
been performed for the case of inelastic scattering. Here we will compare
our work to that of Furnstahl and Wallace [28], where the nucleons in the 
nucleus are described by Dirac bound states.  These authors consider
specific finite nuclei, and point out that the density dependent effects for the elastic
scattering channel are less than those for the reaction channels because
of the identity between initial and final states in the first case. In our paper,
the G-matrix is
constructed for infinite nuclear matter, and no such distinction exists,
namely, we use the same effective interaction for elastic and inelastic
scattering. The implications of this difference in approach will also 
be examined in Section 6. 

Since the primary purpose of this paper is to examine the treatment of 
conventional medium effects, we will compare our predictions in Section 7
to measurements of the cross section and analyzing power for natural-parity,
isoscalar (p,p$'$) transitions where these effects are largest. We have 
chosen the data of Seifert [29] on $^{16}$O and $^{40}$Ca (200~MeV) and 
Chen [30] on $^{28}$Si (180~MeV) since these are the same energies and 
targets that will appear in a future consideration of exotic medium effects
for polarization transfer measurements.  For natural parity transitions,
polarization transfer measurements do not add new information to the
analysis beyond that contained in the cross section and analyzing power
[31] and will not be considered here.
On the basis of these comparisons
we will then be able to evaluate the theoretical tools presented here
for dealing with the nuclear medium.

\vskip.4in
\noindent{\fta 2.~~The nucleon-nucleon interaction}
\bigskip
The starting point for a microscopic calculation of (p,p$'$)
reactions is a realistic free-space NN
interaction which reproduces well NN scattering
observables.  Below pion production threshold, we would expect
that a one-boson-exchange basis would prove adequate.             
This is a very popular and quantitative reduction of a more 
comprehensive model which 
would contain multiple meson-exchange diagrams [22].

In this work, we will use an updated version of the Bonn-B
potential [22].  We have chosen pseudovector coupling for the
pion, and will solve for the NN interaction through the use of the
Thompson equation.  The major bosons in the model are: (a) the
pseudoscalar pion, which is the lightest and hence longest ranged
of the mesons and which provides most of the attractive tensor
force, (b) the $\rho$ vector meson, a 2-pion P-wave resonance
that provides a short-range repulsive tensor force, (c)
the $\omega$ vector meson, a 3-pion resonance that creates a
strong, repulsive central force at short range and that 
contributes to the spin-orbit force, and (d) the fictitious
isoscalar-scalar $\sigma$ meson that represents 2-pion S-wave
exchange and that contributes to the medium range attraction
necessary to bind the nucleus.

These mesons are coupled to the nucleon through the Lagrangians
for the pseudovector ($pv$), scalar ($s$), and vector ($v$)
fields: $${\cal L}_{pv}=-{f_{ps}\over m_{ps}}\overline\psi
\gamma^5\gamma^\mu\psi\partial_\mu\phi^{(ps)}\eqno{(1)}$$
$${\cal L}_{s}=g_{s}\overline\psi\psi\phi^{(s)}\eqno{(2)}$$
$${\cal L}_{v}=-g_v\overline\psi\gamma^\mu\psi\phi_\mu^{(v)}
-{f_v\over 4m}\overline\psi\sigma^{\mu\nu}\psi (\partial_\mu
^{}\phi_\nu^{(v)}-\partial_\nu^{}\phi_\mu^{(v)})\eqno{(3)}$$
with $\psi$ the nucleon and $\phi_{(\mu )}^{(\alpha )}$ the
meson fields (notation and conventions as in Ref.~[22]).  For
isovector (${\rm isospin}=1$) mesons such as the $\pi$ and 
$\rho$, $\phi^{(\alpha )}$ is to be replaced by $\tau\kern-2pt
\cdot\kern-2pt\phi^{(\alpha )}$ with $\tau$ the usual Pauli 
matrices acting in isospin space.

The OBE amplitudes are derived from the Lagrangians in Eqs.~(1)
through (3).  For example, the contribution to the potential
from the isoscalar-scalar $\sigma$ meson is $$\langle{\bf q'}
\lambda_1'\lambda_2'|V_s^{OBE}|{\bf q}\lambda_1\lambda_2\rangle
=-g_s^2\overline u({\bf q'},\lambda_1')u({\bf q},\lambda_1)
\overline u(-{\bf q'},\lambda_2')u(-{\bf q},\lambda_2 )
{({\cal F}_s[({\bf q'-q})^2])^2\over ({\bf q'-q})^2+m_s^2}
\eqno{(4)}$$ where $\lambda_i$ $(\lambda_i')$, with $i=1,2$,
denotes the helicity of the incoming (outgoing) nucleons.
${\cal F}[({\bf q'-q})^2]$ is a monopole form factor that
simulates the short-range physics governed by quark-gluon
dynamics, $${\cal F}_s[({\bf q'-q})^2]={\Lambda_s^2-m_s^2\over
\Lambda_s^2+({\bf q'-q})^2}\ ,\eqno{(5)}$$ with $\Lambda_s$
the cutoff mass of the isoscalar-scalar meson.  The Dirac
spinors are normalized covariantly to $$\overline u({\bf q},
\lambda )u({\bf q},\lambda )=1\ .\eqno{(6)}$$  The OBE
amplitudes for other mesons are given in Refs.~[22,32].  The OBE
potential is the sum of the OBE amplitudes for all the 
exchanged mesons.

Since the parameters of the original Bonn-B interaction were
determined about a decade ago, there has been substantial
progress in both high-precision measurements of NN scattering
and phase shift analyses.  Since a good fit to the NN data
is a crucial prerequisite for an evaluation of the quality of
the (p,p$'$) calculations, the parameters of the model have
been adjusted to bring the predictions into better
accord with current phase shift analyses from the Nijmegem
group up to 325~MeV [33]. 
The $\sigma$ meson is different in the two isospin
channels. As a consequence of this extra degree of freedom, the 
other scalar meson, the $\delta$ meson, becomes unimportant and 
is therefore suppressed. The values 
for the masses, coupling constants, and cutoff parameters for
these mesons are given in Table~1.  This scheme is not fully
satisfactory in the $T=1$ channel.  So in a manner similar to
that used in the construction of the (charge-dependent)
CD-Bonn interaction [34], adjustments have been made to the
$\sigma$-meson coupling constant for some individual partial
waves.  These altered values are given in Table~2.
\topinsert
\centerline{Table 1.~~OBE meson properties}
$$\vbox{\halign{\hfil #\hfil &&\quad\hfil #\hfil\cr
meson& mass (MeV)& coupling $(g^2/4\pi)$& cutoff (MeV)\cr
\noalign{\smallskip}
\noalign{\hrule}
\noalign{\medskip}
$\pi$& 138.03& 13.8& 1700\cr
$\eta$& 548.8& 3& 1500\cr
$\rho$& 769& 0.99 ($f_\rho /g_\rho =6.1$)& 1300\cr
$\omega$& 782.6& 22& 1500\cr
$\sigma\ (T=0)$& 550& 5.966& 2000\cr
$\sigma\ (T=1)$& 550& 8.5& 2000\cr
\noalign{\medskip}
\noalign{\hrule}}}$$
\vskip.4in
\centerline{Table 2.~~$\sigma$ meson couplings}
$$\vbox{\halign{\hfil #\hfil &&\quad\hfil #\hfil\cr
partial wave& coupling $(g^2/4\pi )$\cr
\noalign{\smallskip}
\noalign{\hrule}
\noalign{\medskip}
$^3{\rm P}_1$& 8.9\cr
$^1{\rm D}_2$& 10.05\cr
$^3{\rm P}_2$& 9.11\cr
$^3{\rm F}_3$& 13.5\cr
\noalign{\medskip}
\noalign{\hrule}}}$$
\endinsert
Generally, the agreement with NN data around 200 MeV was found to 
be excellent. 
\vskip.4in
\noindent{\fta 3.~~The G-matrix calculation}
\bigskip
{\bf 3a.~~Conventional Brueckner theory}
\bigskip
It has been almost fifty years since Brueckner and collaborators 
[1] initiated work on a method, further developed by Bethe and 
Goldstone [2,3], to calculate the properties of nuclear matter.                                     
Systematic calculations applying Brueckner's theory started in 
the late 1960's and continued throughout the 1970's [4-6]. For 
a review, see Ref.~[22].

The Brueckner-Bethe-Goldstone model is based on the idea that 
nucleons in nuclear matter move in a single-particle potential 
arising from the interaction with all the other nucleons.  For 
practical reasons, infinite nuclear matter systems are typically 
used in studies of the nuclear many-body problem as a working 
approximation to actual finite systems.  

Kinematics for the two interacting nucleons are defined in terms
of the following momentum vectors.  Consider a nucleon with 
momentum ${\bf k_1}$ colliding with another of momentum 
${\bf k_2}$ embedded in infinite nuclear matter. The Fermi sea 
is defined by the Fermi momentum $k_F$.  If ${\bf k_1}$ and 
${\bf k_2}$ are the momenta of two nucleons in the nuclear 
matter rest frame, it is convenient to introduce the relative
momentum $${\bf k}={1\over 2}({\bf k_1}-{\bf k_2})\eqno{(7)}$$
and one-half of the center-of-mass momentum $${\bf P}=
{1\over 2}({\bf k_1}+{\bf k_2})\ .\eqno{(8)}$$  Conversely, we 
have $${\bf k_{1,2}}={\bf P}\pm {\bf k}\ .\eqno{(9)}$$

The effective two-nucleon interaction in infinite nuclear matter, or 
G-matrix, is a solution of the                                
Bethe-Goldstone equation                      
$$\langle {\bf k}|{\rm G}({\bf P},k_F)|{\bf k_0}\rangle =
\langle {\bf k}|V|{\bf k_0}\rangle -\int{d^3k'\over (2\pi)^3}
{\langle {\bf k}|V|{\bf k'}\rangle Q({\bf k'},{\bf P},k_F)
\langle {\bf k'}|{\rm G}({\bf P},k_F)|{\bf k_0}\rangle\over
E({\bf P},{\bf k'})-E_0-i\epsilon}\eqno{(10)}$$ with ${\bf 
k_0}$, ${\bf k}$, and ${\bf k'}$ the initial, final, and 
intermediate relative momenta, respectively [defined in Eq.~(7)], 
and $V$ is the NN OBE potential defined in the previous section.  
$E$ is the energy of the two-nucleon system, and $E_0$ is the 
same quantity on-shell. Thus $$E({\bf P},{\bf k})=e({\bf P}+
{\bf k})+e({\bf P}-{\bf k})\eqno{(11)}$$ with $e$ the 
single-particle energy in nuclear matter.

Eq.~(10) is density-dependent due to the presence of the Pauli 
projection operator $Q$, defined by $$Q({\bf k},{\bf P},k_F)=
\cases{1,&if $k_{1,2}>k_F$\cr 0,&otherwise\cr}\eqno{(12)}$$
with  $k_{1,2}$ the magnitude of the momenta of Eq.~(9). 
$Q$ prevents scattering into occupied intermediate states.    

To make the calculation more tractable, $Q({\bf k},{\bf P},
k_F)$ is usually replaced by its angle average,  
$${\bar Q}(k,P,k_{F})={\int d\Omega Q({\bf k},
{\bf P},k_F)\over\int d\Omega}\eqno{(13)}$$ with $d\Omega$ the 
solid angle element.  The angle-averaged Pauli function 
approaches the exact value only when the center-of-mass momentum, 
$P$, in Eq.~(8), approaches zero.  In the context of nuclear 
matter saturation (negative incident energies), it has been 
shown that the angle averaging is a good approximation [35].
It has also been demonstrated that the quality of this 
approximation is still good for positive energies up to about 
300~MeV and normal nuclear matter densities [36].       

In nuclear matter, the energy of a single particle with momentum 
$p$, which appears in the energy-denominator of Eq.~(10), is also 
density-dependent and defined by $$e(p)=T(p)+U(p)\ ,\eqno{(14)}$$   
where $T(p)$ is the kinetic energy and $U(p)$ is a single-particle 
potential generated by the average interaction  of all the  
nucleons in the Fermi sea.  For nucleons below and above the Fermi level
we define $$U(p)=\langle p|U|p\rangle =Re\sum_{q\leq k_F} 
\langle pq|{\rm G}|pq-qp\rangle\eqno{(15)}$$ with $|p\rangle$ and 
$|q\rangle$ single particle momentum, spin, and isospin states.

Thus, the propagator in Eq.~(10) depends on $U(p)$ through the 
single-particle energy $\epsilon$ in Eq.~(14). Consequently, 
the determination of G depends on the choice of $U(p)$.  Since the 
potential $U(p)$ must be determined from the reaction matrix 
through Eq.~(15), which depends on G, a solution for both
quantities must be reached in a self-consistent way.

For reasons of numerical convenience, we will use the {\it 
effective mass} ansatz [5], which is equivalent to replacing
the single-particle energy with $${p^2\over 2m}+U(p)
\rightarrow{p^2\over 2m^*}+U_0\ .\eqno{(16)}$$  This way, the 
potential $U(p)$ is parametrized in terms of an effective mass 
$m^*$ and a constant $U_0$.  Eq.~(16) implies $$U(p)={1\over
2}{m - m^*\over mm^*}p^{2}+U_0\ ,\eqno{(17)}$$ that is, the 
potential $U(p)$ has been fitted with a quadratic function of $p$.
Starting from some initial values for $m^*$ and $U_0$, we 
proceed to calculate G from Eq.~(10). From the resulting G, the 
potential $U(p)$ is then calculated via Eq.~(15) and parametrized 
in terms of a new set of $m^*$ and $U_0$. The procedure is 
repeated until the values of $m^*$ and $U_0$ have converged.

The on-shell G-matrix is a set of complex numbers associated
with each of the partial wave quantum numbers usually
used for NN phase shifts.  Since we are below pion production
threshold, these can be checked for agreement with unitarity.
Agreement is excellent at zero density, and represents one
verification of the numerical accuracy of the program.  However
unitarity, particularly for the lower partial waves, is not
obeyed as the nuclear density increases.  In some cases, the
inelasticity parameter differed from one by as much as 18\%
at full nuclear density.  Respecting these violations of 
unitarity in the subsequent transformation to an effective
NN interaction is crucial, especially for the imaginary terms.
Without that, even qualitative agreement with previous
density-dependent analyses would have been impossible.

Finally, from the self-consistently produced G-matrix,              
scattering parameters (positive energies) or the 
bound state properties (negative energies) can be predicted. 
However, nuclear matter calculations based on this approach        
have been only partially successfull, failing to predict 
correctly nuclear matter saturation.  Typically the saturation 
density is too high for reasonable binding energies.  Different 
calculations that change some aspect of the model or use another
NN potential give somewhat different predictions, but all of 
them fail to reproduce simultaneously the correct combination
of saturation energy and density [22].  Since the calculation
at this stage resembles other work on nuclear matter effects,
we will use it later to establish our method and refer to it
as the Brueckner-Hartree-Fock (BHF) theory.  One aspect that
is not included in this framework is an explicit consideration
of the effects of the lower components when the nucleon is
treated as a Dirac spinor.  This extended calculation, which
we will refer to as the Dirac-Brueckner-Hartree-Fock (DBHF)
theory, is described in the next subsection.
\bigskip
{\bf 3b.~~The relativistic approach}
\bigskip
In the early 1970's, a relativistic approach to nuclear structure 
was developed by Miller and Green [37] to study the ground states 
of nuclei.  This was successful in explaining the single-particle 
energy levels, and in particular the spin-orbit splitting.  In 
1974, Walecka [38] published the Dirac-Hartree model for highly 
condensed matter.  Later, Clark and collaborators [39] applied a 
Dirac equation containing a scalar and the time-like component
of a vector field to proton-nucleus scattering.  The Dirac-Hartree 
and Dirac-Hartree-Fock approach to nuclear matter and finite 
nuclei were further developed by Brockmann [40], Horowitz and 
Serot [41], and Serot and Walecka [42].  Finally, a relativistic
extension of Brueckner theory was proposed by Shakin {\it et al.} 
[7], and further developed by Brockmann and Machleidt [8], and 
Horowitz and Serot [9]. Extensive applications and a review
have been presented by ter Haar and Malfliet [10].

A nucleon in the nuclear medium can be viewed as a `bare' nucleon 
that is `dressed' as a consequence of its effective two-body 
interactions with the other nucleons.  Then, the bare and the 
dressed nucleon propagators are related through the Dyson
equation $${\cal G}(p)={\cal G}^0(p)+{\cal G}^0(p)U(p){\cal G}(p)
\eqno{(18)}$$ where $U(p)$ is the relativistic self-energy. Eq.~(18)
has the formal solution $${\cal G}(p)=(\not\kern-2pt p-m-U(p))
^{-1}\ .\eqno{(19)}$$  To be consistent with symmetry requirements, 
the self-energy must have the general Lorentz structure [43]
$$U(p)=U_S(p)+\gamma_0U^0_V(p)-\hbox{\ftb\char'15}\kern-2pt
\cdot\kern-2pt {\bf p}U_V(p)\eqno{(20)}$$ where $U_S(p)$ and $U_V(p)$ 
are an attractive scalar field and a repulsive vector field, 
respectively, and $U^0_V(p)$ is the timelike component of the vector 
field.  It has been shown [43] that $U_V(p)$ is much smaller than 
either $U_S(p)$ and $U_V^0(p)$.  Thus we can write $$U(p)\approx U_S(p)
+\gamma^0U_V^0(p)\ .\eqno{(21)}$$  Now, substituting Eq.~(21) into 
Eq.~(19) leads to the following definitions: $$m^*(p)=m+U_S(p)      
\eqno{(22)}$$ and $$(p^0)^*=p^0-U^0_V(p)\ ,\eqno{(23)}$$ which 
enables us to write the nucleon propagator in nuclear matter, 
Eq.~(19), as $${\cal G}(p)=(\not\kern-2pt p
\kern2pt^*-m^*)^{-1}\ .\eqno{(24)}$$
Thus the Dirac equation in nuclear matter $$(\not\kern-2pt p-m-U(p))
u({\bf p},s)=0\eqno{(25)}$$ can be written simply as $$(\not\kern-2pt 
p\kern2pt^*-m^*)u({\bf p},s)=0\eqno{(26)}$$ with the positive energy 
solution $$u({\bf p},s)=\left({\epsilon_p^*+m^*\over 2m^*}
\right)^{1/2}\left( \matrix{{\bf 1}\cr{\hbox{\ftc\char'33}
\cdot{\bf p}\over\epsilon^*_p+m^*}\cr}\right) \chi_s
\eqno{(27)}$$ where $\chi_s$ is a Pauli spinor, and $$
\epsilon_p^*=(m^{*2}+{\bf p}^2)^{1/2}\ .\eqno{(28)}$$ 
$u({\bf p},s)$ is formally identical to a free-space spinor, but 
with $m$ replaced by $m^*$. 

As mentioned earlier, Dirac spinors use the covariant normalization
$u{\bar u}=1$.  However physical nucleon states must be normalized 
by $ww^\dagger =1$, where $w=\sqrt{m^*/\epsilon^*}\times u$. 

From Eq.~(21), the single-particle potential is $$U(p)={m^*\over
\epsilon_p^*}\langle p|U_S+\gamma^0U_V^0|p\rangle\eqno{(29)}$$
where $|p\rangle$ is a Dirac spinor as in Eq.~(27) and $\langle 
p|$ the adjoint spinor.  In analogy with the usual Hartree-Fock 
definition, $$U(p)=Re\sum_{q\leq k_F}{m^{*2}\over\epsilon^*_p 
\epsilon^*_q}\langle pq|{\rm G}|pq-qp\rangle\ ,\eqno{(30)}$$ which is 
similar to the (non-relativistic) Brueckner-Hartree-Fock 
definition of the single-particle potential in Eq.~(15).  This
situation is now technically identical to the one encountered in 
the non-relativistic case. Eq.~(30) and the relativistic (Thompson form)
G-matrix equation               
(with the potential $V$ written in terms of the in-medium spinor
from Eq.~(27)) must be solved self-consistently for G and $U(p)$. 
$U_S(p)$ and $U_V^0(p)$ are momentum dependent, but to a good 
approximation they can be taken as constant [7,9]. Thus, the 
single particle potential becomes $$U(p)={m^*\over\epsilon_p^*}
U_S+U_V^0\ .\eqno{(31)}$$  Eq.~(31) is a consequence of Eq.~(29) 
together with our normalization conventions: $u{\bar u}=1$ and 
$uu^{\dagger}=\epsilon_p^*/m^*$, with ${\bar u}=u^{\dagger} 
\gamma^0$.  In analogy with Eq.~(17), Eq.~(31) parametrizes 
the energy of a particle in nuclear matter in terms of two 
constants, which is clearly equivalent to the effective mass 
ansatz used in the previous subsection for the non-relativistic 
case.

The main difference between BHF and DBHF is that for DBHF the 
nucleon wave function, Eq.~(27), is obtained self-consistently, 
while in BHF the free-space solution is used.  Such a difference 
turns out to be of fundamental importance.  As a consequence of 
the reduced nucleon mass, the lower component of the nucleon 
spinor is larger than in free space.  This is well known to 
produce a density-dependent, repulsive many-body effect, with 
the result that saturation can be predicted at the correct 
density [7].  To demonstrate how this happens, consider a 
vertex with a scalar coupling, such as the one involved in 
$\sigma$-exchange. Using Eq.~(27), we have $${\bar u}({\bf p'},
s')u({\bf p},s)\propto\chi_s'\left( 1-\left({\hbox{\ftb\char'33}
\cdot{\bf p}\over\epsilon_p^*+m^*}\right) \left({\hbox{\ftb
\char'33}\cdot
{\bf p'}\over\epsilon_p'^*+m^*}\right) \right) \chi_s\ ,
\eqno{(32)}$$ which is one of the two vertices involved in the 
(attractive) $\sigma$-exchange potential diagram of Eq.~(4).
As the nucleon mass decreases with higher densities, the 
increased lower component of the Dirac spinor will quench this              
attractive contribution.  Thus we have a repulsive effect. 
In contract, consider a vertex with a vector coupling, such 
as the one involved in the (repulsive) $\omega$-exchange. The 
time component of such vertex gives $${\bar u}({\bf p'},s')
\gamma_0u({\bf p},s)\propto\chi_s'\left( 1+\left({\hbox{\ftb
\char'33}
\cdot{\bf p}\over\epsilon_p^*+m^*}\right) \left({\hbox{\ftb
\char'33} 
\cdot{\bf p'}\over\epsilon_p'^*+m^*}\right) \right) \chi_s\ ,
\eqno{(33)}$$ which is enhanced as the nucleon mass decreases,
resulting in additional repulsion.  This argument also explains 
the typical relativistic enhancement of the spin-orbit force, 
which is mainly provided by $\omega$ exchange. 

Another consequence of the larger size of the spinor lower 
component at higher densities is that the usual pseudoscalar 
($\gamma_5$) coupling for the pion cannot be used in this 
context, as it would produce unreasonable results.  This can be 
understood by noticing how the pseudoscalar vertex $${\bar u}
({\bf p'},s')i\gamma_5u({\bf p},s)\propto\chi_s'\left({\hbox{\ftb
\char'33}\cdot{\bf p}\over\epsilon_p^*+m^*}-{\hbox{\ftb\char'33}
\cdot
{\bf p'}\over\epsilon_p'^*+m^*}\right) \chi_s\eqno{(34)}$$
couples lower and upper components of the Dirac spinors. 
As the nucleon mass decreases (to approximately 70\% of its 
free-space value at nuclear matter density), this contribution 
(and thus the entire $\pi$-exchange) grows to unreasonably 
large values.  The pseudovector coupling of Eq.~(1) does not 
cause these problems, which is the reason why we choose it 
here for the coupling of the pion to the nucleon.

This subsection has summarized the formalism for a relativistic 
approach to nuclear structure which is based on a relativistic 
extension of Brueckner theory.  Relativistic meson-exchange 
potentials are appropriate for this framework, with the 
pseudovector coupling for the $\pi$NN vertex.  The nucleon 
self-energy can be parametrized in terms of an attractive 
scalar and a repulsive vector field.  Together with the single 
particle wave function, these are determined self-consistently. 
The attractive scalar field leads to a reduction of the nucleon 
mass; this effect increases with density and facilitates 
the prediction of nuclear saturation at the 
correct energy and density. 
\bigskip
{\bf 3c.~~Medium effects on the nuclear force}
\bigskip
We conclude this section by demonstrating the effects of the medium 
(Pauli/binding and relativistic) on the nuclear force using as
examples some selected partial waves.

In Figs.~1 and 2, we show the real and imaginary parts of the 
t- or G-matrix for some chosen S and P partial waves.  In all 
cases, the solid curve is the free-space calculation (t-matrix), 
the short-dashed curve is the Brueckner-Hartree-Fock 
G-matrix (BHF), and the 
long-dashed is the relativistic G-matrix (DBHF). Plotted are 
half-off-shell t- or G-matrix elements as a function of 
momentum.  At a bombarding energy of 200~MeV, the on-shell 
momentum is 1.55~fm$^{-1}$.  The Fermi momentum is fixed at 
1.0~fm$^{-1}$. 
\topinsert
\vbox to 4.6 truein {\vss
\hbox to 6.5 truein {\includegraphics{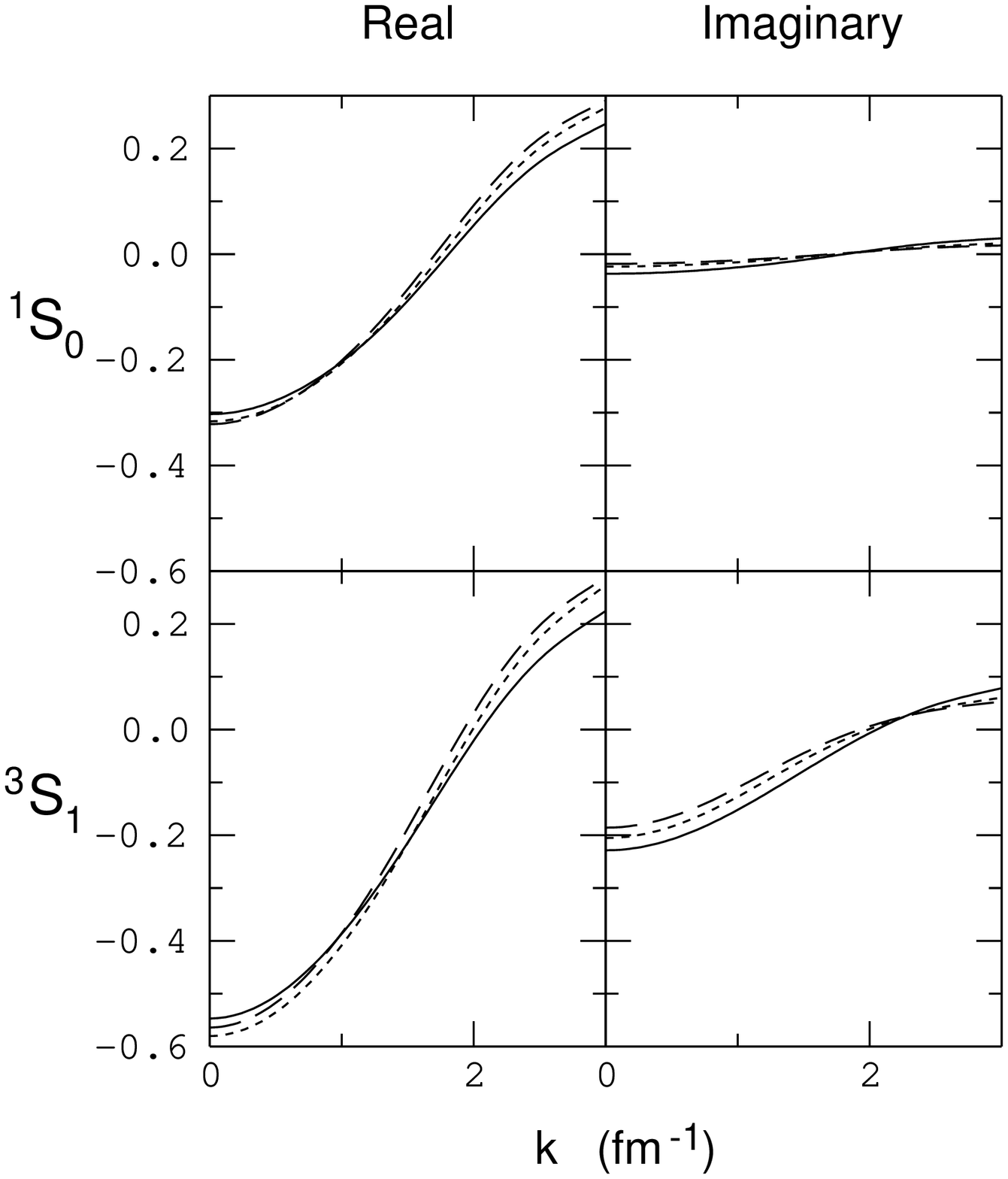}\hss}}
\noindent{\it Figure 1.}~~The real (a) and imaginary (b) parts
of the half-off-shell t- or G-matrix for the $^1$S$_0$ and
$^3$S$_1$ partial waves at 200~MeV.  The on-shell momentum is
1.55~fm$^{-1}$.  The solid line is the free space calculation.
The short-dashed and long-dashed lines are obtained with BHF and DBHF
calculations, respectively.  The Fermi momentum is 1.0~fm$^{-1}$.
\endinsert
\topinsert
\vbox to 4.6 truein {\vss
\hbox to 6.5 truein {\includegraphics{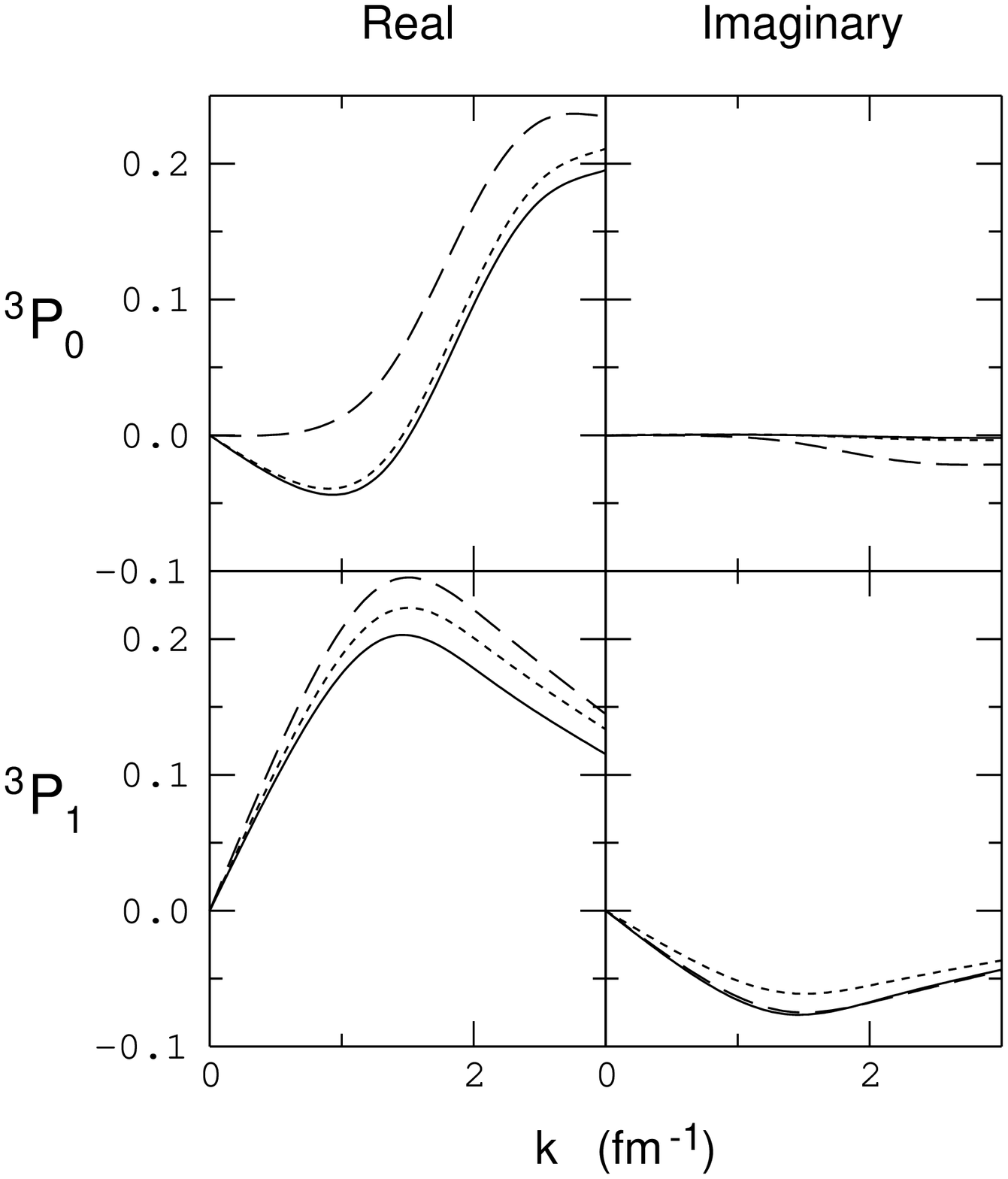}\hss}}
\noindent{\it Figure 2.}~~The same as Fig.~1 for the $^3$P$_0$
and $^3$P$_1$ partial waves.
\endinsert

For the $^1$S$_0$ and $^3$S$_1$ partial waves in Fig.~1 and values 
of the momenta near the on-shell point (which are the most 
relevant for the processes we are considering) the t-matrix 
elements (solid lines) are negative.  BHF medium 
effects are then observed to move the curves upward toward 
smaller absolute values.  The same medium effects, 
when applied to the case of negative energies (that is, in 
nuclear matter binding energy calculations), produce repulsion.
The trend is rather similar in both S-waves.    

For the $^3$P$_0$ and $^3$P$_1$ partial waves in Fig.~2, BHF 
medium effects also move the curves upward (compare solid and 
short-dashed lines). 
In nuclear matter binding calculations [22] the 
medium effects on the P-waves are repulsive and generally small.

The change to a relativistically-based G-matrix calculation
generally provides extra repulsion, as expected.  In the S-waves, 
this comes primarily from the suppression of the $\sigma$-meson, as 
discusseded in the previous subsection.  For the P-waves, 
relativistic effects are much more dramatic.  Besides the 
suppression from the $\sigma$, they are associated with the
enhancement of the spin-orbit force, a term in the NN interaction
to which the P-waves are especially sensitive. 

\topinsert
\vbox to 3.6 truein {\vss
\hbox to 6.5 truein {\includegraphics{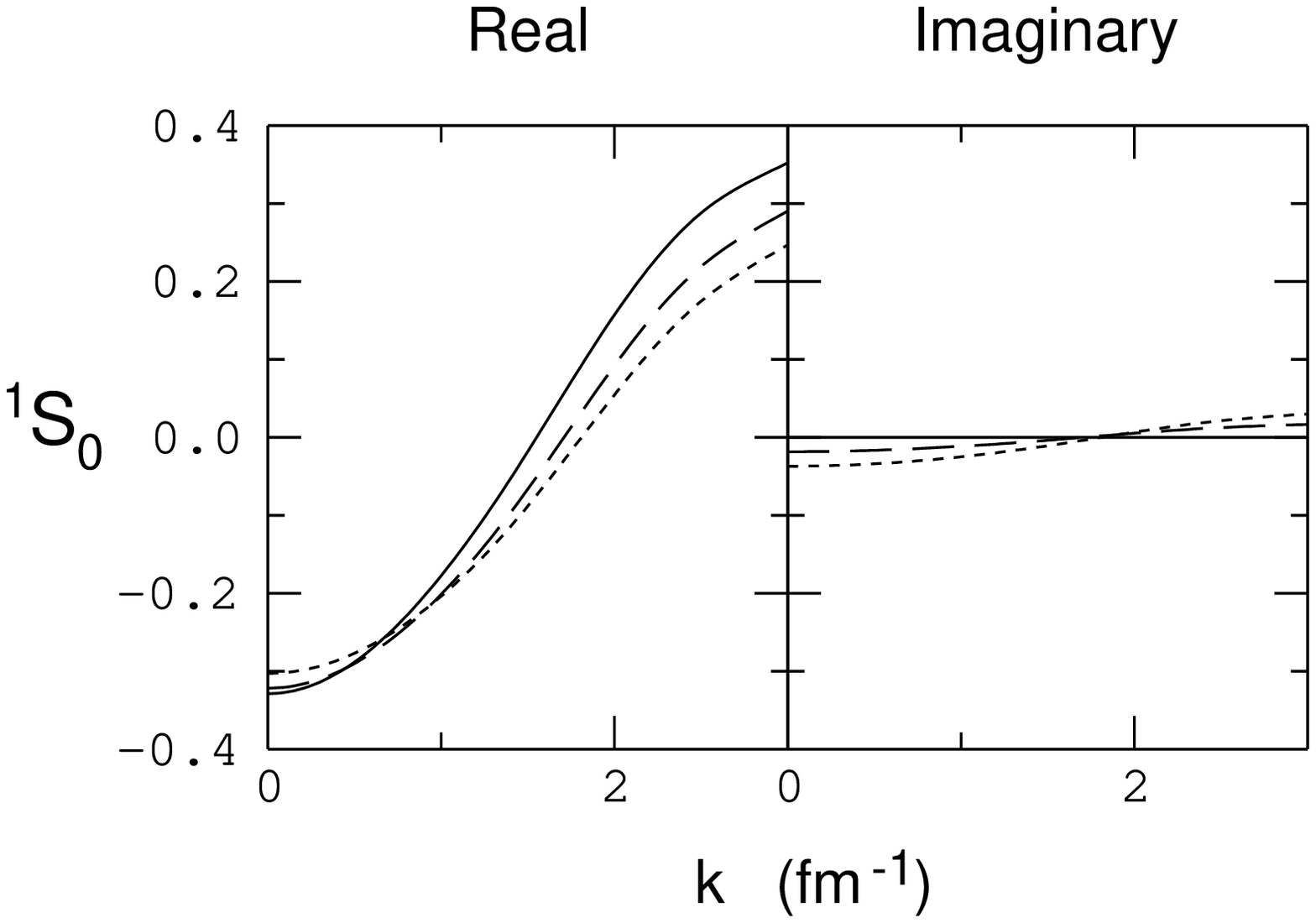}\hss}}
\noindent{\it Figure 3.}~~Density dependence of the $^1$S$_0$
half-off-shell partial wave amplitude at 200~MeV.  The solid 
curve is for $k_F=0$~fm$^{-1}$, the short-dashed for $k_F=1.0$~fm$^{-1}$,
and the long-dashed for $k_F=1.35$~fm$^{-1}$.
\endinsert
Overall, medium effects can be large.  The density dependence
is shown in Fig.~3 for one case only.  A proper treatment of the 
density-dependence of the effective NN interaction 
is crucial for reliable nuclear structure and reaction calculations,
as we will explore in the rest of this paper. 
\vskip.4in
\noindent{\fta 4.~~Transformation to an effective NN interaction}
\bigskip
In its simplest form, the DWBA requires a perturbing potential
which, for (p,p$'$) reactions, is the NN potential described in 
Sect.~2.  For practical calculations it has proven better to
replace this with the t- or G-matrix, thereby including the full
complexity of multiple meson exchange and creating what is known
as the Distorted Wave Impulse Approximation (DWIA) [44].
In the case of the strong force, this also has the advantage of
reducing the size of the interaction, since short-range repulsive
meson exchanges tend to cancel the effects of the longer-range
attractive parts.  This makes the perturbation theory more 
tractible, since in many cases a first-order calculation suffices,
and that calculation is much more accurate.

The realization of such calculations is traditionally performed by 
coordinate-space computer programs for which the t- or G-matrix
must be transformed from momentum space to coordinate space.
A popular representation for this effective NN interaction has
been a series of Yukawa functions [21] which may
use the same expansion for both coordinate and momentum spaces.
In part, this choice is motivated by the observation that the
potential in Eq.~(10), $\langle{\bf k}|V|{\bf k_0}\rangle$, is
of Yukawa type and, especially for the spin-dependent parts of
the interaction, is the leading contribution to the t- or
G-matrix.  The expansion coefficients are then obtained by making
a fit of the expansion, given a pre-chosen set of ranges for the
Yukawa functions, to the on-shell NN scattering amplitudes.

Some versions of coordinate-space 
DWIA calculations make a separation into
direct and exchange parts by explicitly interchanging the indices
of the projectile and struck nucleon in the exit channel.  This 
results in a six-dimensional coordinate-space integral for the
DWIA scattering matrix.  A motivation for the separation is to
deal explicitly with some (p,p$'$) and (p,n) transitions in which
their unnatural parity implies that there is no contribution to
certain parts of the scattering matrix at zero range. 
Whatever contributions arise, particularly having to do with the
spin dependence of the tensor parts of the interaction, are very
sensitive to the distinct contributions of direct or exchange.
While in our model direct and exchange are in principle distinguishable since
the t- or G-matrix originates from the solution to a potential
problem, historically a number of interactions were created by
reproducing the amplitudes from a phase shift analysis of NN
scattering where such a distinction is not possible.  So in this
expansion it was assumed that the sum over Yukawa functions was
the direct part, and the full NN scattering amplitude was
generated using the symmetries of NN scattering [21].
In effect, this leads to a different set of functions that are
fit to the NN amplitudes.  The hope is that this artificial
separation will then work in a practical calculation.

There are a number of reasons to expect that this approximation is
actually satisfactory.  For the central and spin-orbit parts of the
effective NN interaction, the spin operator for the direct and
exchange amplitudes is the same.  So in the end, these two pieces
are added together again just as they were when the original fit
was made to the NN scattering amplitude.  The quality of this
approximation is enhanced since the central parts are usually
largest at zero range where the distinction between these two pieces
disappears.  For the tensor terms, it is known that the direct
piece has to depart from zero momentum transfer as $q^2$, so any
NN amplitude near $q=0$ must be part of the exchange piece.  Also,
the parts of the interaction associated with the $S_{12}(\hat q)$
and $S_{12}(\hat Q)$ spin operators [$\hat{\bf q}=({\bf k'}-{\bf k})
/|{\bf k'}-{\bf k}|$,
$\hat{\bf n}=({\bf k}\times{\bf k'})
/|{\bf k}\times{\bf k'}|$, and $\hat{\bf Q}=\hat{\bf q}\times\hat
{\bf n}$]
arise essentially from the interplay of the real potentials from
$\pi$- and $\rho$-meson exchange with little contribution from the
integral of Eq.~(10).  Thus, if the Yukawa functions simulate well
the real parts of the $\pi$- and $\rho$-meson exchanges, other parts
of the tensor interaction will fall into place.
Since our long-term purpose is to bring together diverse sources of
information about the NN interaction, it was decided that we would
continue to make the separation between direct and exchange in this
model-dependent way.

These observations are also useful in assessing where we expect there
to be large density-dependent effects.  Since most of the density
dependence in Eq.~(10) enters through the integral,                  
(this is exactly true for BHF effects, which only
enter the picture in second or higher order), terms in the 
effective NN interaction where $\langle{\bf k}|{\rm G}({\bf P},
k_F)|{\bf k_0}\rangle\approx\langle{\bf k}|V|{\bf k_0}\rangle$
will be relatively free of these effects.  This applies to the
tensor parts of the interaction.  In fact, the largest differences
arise for the isoscalar central and spin-orbit parts of the 
effective NN interaction, and these are the parts that drive the
excitation of the natural parity states we will study.  Since
$\langle{\bf k}|V|{\bf k_0}\rangle$ is real, the imaginary parts
of the effective NN interaction all arise from the 
density-dependent integral, and would be expected to manifest the
effects of the nuclear medium to the extent that they make large
contributions to the total transition amplitude for their
particular spin operator.

The various amplitudes for the effective NN interaction are
separated among the five spin operators according to $$t=t_0+
t_\sigma\sigma_1\kern-2pt\cdot\kern-2pt\sigma_2+t_S(\sigma_1+
\sigma_2)\cdot n+t_{TD}S_{12}(\hat q)+t_{TX}S_{12}
(\hat Q)\ .\eqno{(35)}$$
The five pieces in order are the central, spin-spin, spin-orbit,
tensor direct, and tensor exchange.  Each amplitude is divided
into two pieces that correspond to the isospin operators {\bf 1}
and $\tau_1\kern-2pt\cdot\kern-2pt\tau_2$.  There are then 10 
complex amplitudes all together. Each piece
is represented for the direct part by a sum of from 2 to 4 Yukawa
functions.

The coefficients for each Yukawa function in each sum are determined
so as to best reproduce the NN amplitudes.  From the G-matrix
calculation, there are complex values of the G-matrix elements.
These form an array whose index consists of the partial wave
quantum numbers.  Likewise the Yukawa coefficients can be put into
an array, and there is a linear transformation that connects the
two.  For this transformation, we actually convert the G-matrix
elements into the corresponding S-matrix representation.  The
transformation, discussed more fully in the Appendix, is then the
matrix {\bf M} in ${\bf S}={\bf MV}$ where {\bf S} is a vector
composed of the S-matrix elements and {\bf V} is a vector
of Yukawa coefficients.  Because of the
long range of the pion tail, it is useful to extend the range of
S-matrix elements up through $J=15$.  In this case there are more
partial waves than Yukawa coefficients, and a fit is necessary.
Since the system is linear, this can be accomplished in one step.
Using $dS_i$ as the error on any particular S-matrix element
(set here to the same value for all partial waves),
we can define the vector and matrix $${\cal S}_j=\sum_i{S_iM_{ij}
\over dS^2_i}\qquad {\cal M}_{jk}=\sum_i{M_{ik}M_{ij}\over
dS^2_i}\eqno{(36)}$$ from which the solution for the Yukawa 
coefficients is simply $$V_k=\sum_j
{\cal M}_{jk}^{-1}{\cal S}^{}_j\ .\eqno{(37)}$$
For our work near 200~MeV, we chose the ranges used by Franey
and Love [45] at 210~MeV and found satisfactory results
for all nuclear densities.

Once the transform matrix {\bf M} is available, it may be used
to go from the Yukawa coefficients back to the NN G-matrix
elements.  It is also possible to compare the effective NN 
interaction, for example as a function of momentum transfer,
with the corresponding NN scattering amplitudes.  So a number of
checks of this transformation are possible.  For the polarization
observables, the differences are in all cases well within 0.01,
so the quality of this fit is not an issue for subsequent
comparisons.  It is, in fact, somewhat remarkable that such a
high quality of fit is available with so few terms in the
Yukawa expansion for each part.  The spin-orbit, for example,
contains only two for each of the real and imaginary parts.
This is, perhaps, further support for the
use of Yukawa functions as the direct part of the effective NN
interaction.

This transformation is similar to that used by 
Karataglidis {\it et al.}\ [46] to
provide an effective NN interaction for their study of 
$^{12}$C(p,p$'$) reactions.  In their case, they also consider
the off-shell parts of the G-matrix within about 0.5~fm$^{-1}$
of the on-shell point.  We have used their formulation, and do
not find that this added sensitivity significantly changes the
(p,p$'$) calculations. So we have not incorporated that
extension into the work here.
\vskip.4in
\noindent{\fta 5.~~The DWIA calculations}
\bigskip
In the longer term, we wish to use the techniques set forth here
to examine the question of medium modifications for a large body
of polarization transfer data taken at 200~MeV.  Here, we evaluate
the calculations at this energy using (p,p$'$) measurements where the
conventional medium effects are expected to be large and 
where the theory could be capable
of describing the data.  For this purpose, we have
chosen natural parity transitions measured in $^{16}$O and 
$^{40}$Ca at 200~MeV [29], and in $^{28}$Si at 180~MeV [30].
In the original work where these data were first reported,
measurements of the cross section and 
analyzing power were matched with a phenomenological effective
NN interaction using the DWIA program LEA [23].  We have obtained
a copy of this program and have used it in a manner modelled after 
the applications described by Seifert and Chen.

LEA makes use of a pseudopotential form of the Born approximation
in which each term of the effective NN interaction is folded
with the appropriate piece (depending on spin operator coupling)
of the nuclear density.  The resulting effective potential is 
then used to calculate the transition amplitudes by overlapping
it with a distorted wave description of the projectile motion
and the appropriate piece of the transition density.
This scheme uses a system in which the spatial coordinate is tied
to the position of the projectile.  Thus it is most
appropriate for the calculation of the
direct part of the scattering.  Exchange is handled in a zero
range approximation that adds this piece of the effective NN
interaction to the direct
part, and the overlap is calculated as it would be if the 
outgoing nucleon were the projectile.  For all prior work
with LEA [29,30], the exchange part of the effective NN
interaction has been calculated at
a momentum transfer of zero, then added to the direct
part.  It is thus a 
constant, independent of momentum transfer.

In the investigation of the high spin states of $^{28}$Si, we
modified this treatment of exchange
to allow the exchange momentum transfer $Q$ to vary
with the momentum transfer $q$ [20].  The incoming and outgoing 
proton
momenta were constrained to their laboratory values.  This means
that for a given scattering angle, the outgoing momentum from a
particular state will be different than it would be if the
scattering took place from a nucleon in free space.  To make
this scattering happen with on-shell kinematics, it is necessary
to have the struck nucleon in motion
prior to the collision.  There are an infinite
number of solutions to this problem.  We chose the
one with the smallest possible momentum for the struck nucleon
relative to the center-of-mass motion of the target nucleus,
with the idea that the momentum-space wavefunction for the 
bound nucleons falls off as a function of momentum and that the
typical value would be close to this limit.  This permits the
calculation, as a function of scattering angle, of both the momentum
transfer $q$ and the exchange momentum transfer $Q$, taking the
latter to be the change in momentum when the struck nucleon is the
detected nucleon.                                                 
In Fig.~4 we compare two calculations for
the $^{28}$Si(p,p$')^{28}$Si reaction to the first $2^+$ state.
The solid curve uses the original zero range approximation
[$Q(q)=Q(q=0)$] of LEA, and the heavy dashed line is based on
the functional values of $Q(q)$ just described.  The curves
are essentially indistinguishable.  So while this choice is
important for the treatment of the exchange tensor parts of
the effective NN interaction, it is not a serious issue here.

\topinsert
\vbox to 3.6 truein {\vss
\hbox to 6.5 truein {\includegraphics{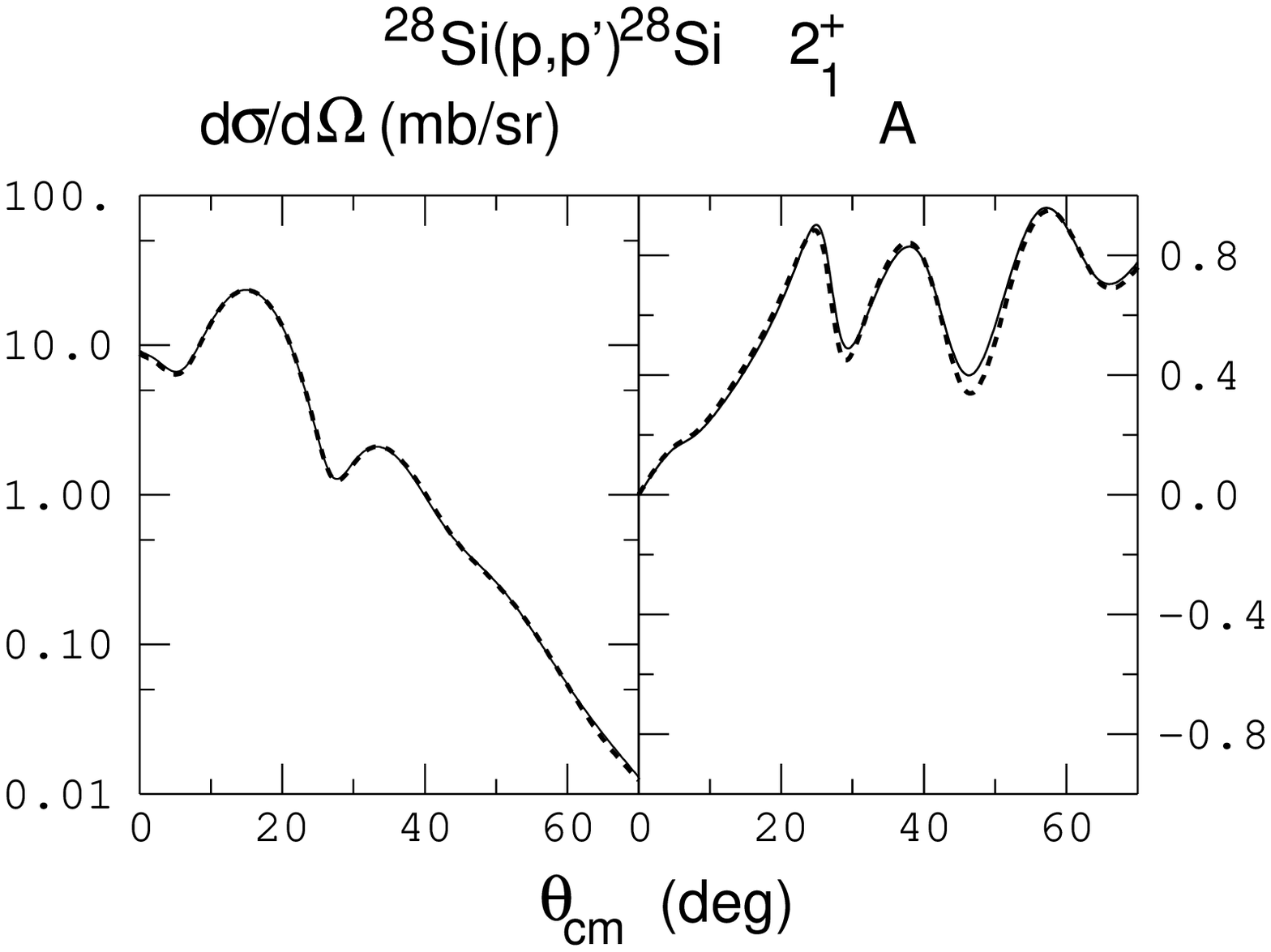}\hss}}
\noindent{\it Figure 4.}~~Calculations of the cross section
and analyzing power for the first $2^+$ state in $^{28}$Si.
The thin solid curves use the zero range prescription of LEA [23],
the heavy dashed curves use the formulation described in the text.
\endinsert

In other aspects, our calculations with the DWIA program LEA
followed the scheme described by Seifert [29] and Chen [30].
The entrance and exit channel distortions were calculated using
a first-order folding prescription (t$\rho$).  The same 
density-dependent interaction is used for both
the elastic and inelastic calculations except for the small
change to the exchange part required by the different reaction
Q-value.  However, there are corrections to the 
density-dependence that are different between the elastic and
the inelastic channels [47,48], and these have been included.
In addition, there is a small downward rescaling of the effective
NN interaction that has to do with the change between the 
nucleon-nucleon and the nucleon-nucleus frames of reference
[49,21], and this has also been included.

The ground state nuclear matter density is needed to determine
the density at which the interaction should be evaluated.  We 
have taken the longitudinal formfactor for elastic electron
scattering [50] and corrected it for the charge distribution
of the proton using an unfolding procedure available in LEA.  
What remains from the deconvolution is assumed
to be a point nucleon density that applies
to both protons and neutrons, since all of the targets
considered here have equal proton and neutron numbers.  This
density is then evaluated at the position of the projectile.
No alterations are made to account for different densities
at the position of the struck nucleon.
This ``local density approximation''
is similar to the zero range approximation for the
exchange part of the interaction, in that both assume that the
strongest part of the transition takes place when the projectile
and struck nucleons overlap.  

The transition formfactors have been calculated by Seifert [29]
and Chen [30] in a series expansion based again on longitudinal
electron scattering measurements.  We use their expansions
here.
\vskip.4in
\noindent{\fta 6.~~Variations within the Effective Mass Ansatz and their Impact}
\bigskip
For both the BHF and DBHF treatments of the in-medium interaction,
the exact nuclear matter potential $U(p)$ of Eqs.~(15) or (30) is 
approximated by an analytical function containing the kinetic energy
term and a constant [see Eqs.~(17) and (31)]. This approximation facilitates
later developments in the self-consistent procedure [5]. The two 
adjustable coefficients are usually obtained by matching the approximating
function to the exact $U(p)$ at two values of $p$. For nuclear matter
calculations, where all momenta are below $k_F$, the usual procedure
[22] is to match at $p=0.7k_{F}$ and $k_{F}$. To illustrate the quality
of this approximation, the solid curve in Fig.~5 shows the exact DBHF
potential of Eq.~(30) for $k_{F}=$1.35~fm$^{-1}$. The approximating
function matched at low values of $p$ is shown by the long dashed curve.
When considering scattering by nucleons in the continuum, much higher
momenta are involved, and we see from Fig.~5 that the long dashed curve
becomes a poor approximation for high momenta. The laboratory beam
momentum at 200 MeV is 3.2 fm$^{-1}$, and is shown by the arrow in 
Fig.~5. The approximation can be improved in the vicinity of the beam
momentum by matching the approximating function at higher momenta.
We have chosen $2k_F$ and $4k_F$, which spans the range of momenta
covered here. This choice gives rise to the short dashed curve, which is
seen to rise more gradually with $p$, and thus correspond to a larger
(and closer to the free value) effective nucleon mass. 
\topinsert
\vbox to 3.6 truein {\vss
\hbox to 6.5 truein {\includegraphics{finfig5.ps}\hss}}
\noindent{\it Figure 5.}~~The momentum dependence of the exact (DBHF)
in-medium NN potential $U(p)$ at $k_{F}=$1.35~fm$^{-1}$ (solid). The 
two approximate potentials are matched at low momentum (long dash)
and high momentum (short dash). The arrow marks the beam momentum
corresponding to a laboratory energy of 200~MeV. 
\endinsert
\topinsert 
\vbox to 3.6 truein {\vss
\hbox to 6.5 truein {\includegraphics{finfig6.ps}\hss}}
\noindent{\it Figure 6.}~~The effective nucleon mass as a function    
of $k_F$. The left and right panels show the masses obtained
using low and high matching points as explained in the text. The 
values for BHF (dashed) and DBHF (solid) are marked. 
\endinsert

Figure 6 shows the effective nucleon mass as a function of the Fermi
momentum for the cases we have just described, namely BHF or DBHF for
both low or high matching momenta. For either matching criterion, the
BHF and DBHF masses are close to each other, with the DBHF masses being 
slightly smaller. The DBHF high momentum match was checked and 
still found
to be consistent with the saturation properties of nuclear matter. (We
obtained saturation at 0.17~nucleons/fm$^{3}$ with a binding energy
of 14.4~MeV.) 

It is useful to compare the impact of each of these G-matrix schemes
on the 
effective NN interaction. For this we have chosen the central, isoscalar,
spin-independent interaction at $q=0$ [first term in Eq.~(35)].
The real part is shown in the left and center panels of Fig.~7 for the 
four cases just discussed. The difference between DBHF and BHF is here
much more dramatic than is apparent from the values of the effective
nucleon mass, primarily because the effective mass in DBHF also alters
the strenght of the terms in the OBE potential. For the standard BHF
treatment, these terms remain unchanged with changing nuclear density.
Clearly, choosing the higher matching momenta reduces the size of both
effects.
The imaginary part of the same effective NN interaction
term (not shown here) is negative in all cases, and becomes weaker   
as the density increases. Also, there is much less variation among
the different cases. 

\topinsert
\vbox to 3.5 truein {\vss
\hbox to 6.5 truein {\includegraphics{finfig7.ps}\hss}}
\noindent{\it Figure 7.}~~Variation with nuclear density of the real isoscalar,
central, spin-independent effective NN interaction at $q=0$.    
The left and center panels repeat the cases of Fig.~6. The right
panel includes cases from the literature: Furnstahl-Wallace for 
inelastic and elastic channels [28], Nakayama
and Love [26], and the empirical effective NN interaction of H. 
Seifert [29]. 
\endinsert
\topinsert
\vbox to 3.6 truein {\vss
\hbox to 6.5 truein {\includegraphics{finfig8.ps}\hss}}
\noindent{\it Figure 8.}~~Real central potentials calculated using the folding
model for $^{40}$Ca at 200~MeV for 
some of the interactions of Fig.~7.
\endinsert
The size of the interactions shown in Fig.~7 will also have an effect
when used in a folding model to produce an optical potential for Distorted
Wave calculations. Given a point nucleon density distribution for a nucleus,
this model averages the interaction across the nucleus for each radial
point in the optical potential. 
Values of the interaction term shown in Fig.~7 
for low density contribute to an 
attractive real central optical potential, and this will be typical of the 
folded optical potential in the low-density region of the nuclear surface.
As the local density increases, the stronger medium effects associated
with the DBHF G-matrix will push the optical potential positive. The 
$^{40}$Ca optical potentials for 200 MeV corresponding to some of the 
curves of Fig.~7 are shown in Fig.~8, and the resulting predictions for 
differential cross section and analyzing power are compared to 
elastic scattering data on $^{40}$Ca [29] in Fig.~9.

In comparing potentials, it is important to realize that the most
important region is near the nuclear surface from 3 to 5 fm.
The free case (duplicated for left and center panels) uses only the 
free space interaction, and produces a potential that is too
attractive. This attraction 
tends to draw elastic scattering flux into the nucleus
where the spin-orbit potential becomes more efficient at separating 
spin up and down components in the proton beam. This produces an
analyzing power that is too positive. Obtaining the right amount of surface
attraction, and the right average value for the analyzing power, then becomes
one important criterion for evaluating the folded optical potential.

It is interesting to observe that the BHF calculation with low momentum
matching points is free of these problems. In fact, as can be seen in the
lower left panel of Fig.~9, the dashed curve represents an excellent
prediction for the analyzing power. While too large at angles past 30$^\circ$,
the cross section in the upper left panel of Fig.~9 is also an improvement.
The success of this model has been noted on many occasions, and much 
effort has gone into a microscopic treatment of elastic scattering
[51--56] and its extension into a momentum space representation
[57--59].  For lower
mass target, the small rise in the BHF optical potential near 3~fm grows
until it is near 0~MeV.  This produced in the early 1980's work on the 
``double humped" potential for $^{12}$C [60]. Angular distributions at
very large scattering angles tended to confirm this prediction in 
phenomenological potential shapes. At the same time, the imaginary part of
the free interaction was too absorptive. Moving to a BHF treatment cut the 
size of this almost in half, which was important in obtaining consistency
with other calculations of the mean free path of protons in nuclei [61,62].
\topinsert
\vbox to 3.9 truein {\vss
\hbox to 6.5 truein {\includegraphics{finfig9.ps}\hss}}
\noindent{\it Figure 9.}~~Elastic scattering cross section and analyzing power
measurements for $^{40}$Ca at 200~MeV.  The calculations use the
optical potentials of Fig.~8.
\endinsert

It is interesting to note in passing that the same time produced a number
of studies of elastic proton scattering based on the Dirac equation
and large scalar and vector optical potentials. These also did well
in reproducing data, and when reduced to a Schr\"odinger form potential
as in Fig.~8, a ``double hump" form was also produced. Thus there appeared to
be two explanations of the same phenomenological result. 
 
Returning to our G-matrix schemes, 
from the left panels of both Fig.~8 and 9 it would appear
natural to conclude that the DBHF calculation overestimates medium 
effects, while BHF produces a more reasonable description of both
cross section and analyzing power. On the other hand, moving to the 
center panel (high momentum match) shows that the DBHF predictions
improve noticeably. Thus different choices within the effective mass
approximation seem to have 
considerable impact on the size of the medium effects, which will be reflected
on the distorting potential and scattering observables. 

At this point the results of Furnstahl and Wallace [28] are instructive.
They start from an OBE potential that is fully relativistic [63], and 
solve the Dirac equation in the nucleus. They also include Pauli 
blocking and nuclear binding effects. The interactions from their work
suitable for elastic and inelastic scattering are shown, together with
others, in the right panels of Figs.~7-9. We notice that the inelastic
interaction of Furnstahl and Wallace has the same problem as our 
DBHF when compared with the elastic scattering data, namely a cross
section that is too large at large scattering angles and an analyzing
power that is too negative.

By treating finite nuclear systems, Furnsthal and Wallace point out
that in the elastic scattering channel, the identity between the entrance
and exit channels removes one power of the M\"oller operator, thus 
significantly reducing the relativistic effects. These calculations are
shown by the solid lines in the right hand panels of Figs.~7-9.
For this case, both cross section and analyzing power are in better   
agreement, though their behaviour at large scattering angles suggests that
medium effects in this model are still too large, an observation that
echoes their comparison to (p,p$'$) measurements [64]. 

The above comparison suggests that a reduction in the Dirac effects of
our model by about the size of the M\"oller operator change in Furnsthal
and Wallace would be likely to improve the agreement with the elastic
scattering data. Thus the ingredients described here would appear to be
adequate to understand proton elastic scattering within a G-matrix approach
that also reproduces the density and binding energy of nuclear matter.
\topinsert
\vbox to 3.6 truein {\vss
\hbox to 6.5 truein {\includegraphics{finfig10.ps}\hss}}
\noindent{\it Figure 10.}~~Measurements of the cross section and analyzing power
for the first $3^-$ state in $^{40}$Ca.  The curves are described in
Fig.~7
\endinsert

Since the primary focus of this paper is to establish a reliable basis
for inelastic scattering calculations, it is important to examine the
implications for this process of the model variations described above.
In Fig.~10 we present calculations for the $^{40}$Ca(p,p$'$)$^{40}$Ca
reaction to the 3$^{-}$ state at 3.736 MeV. For the left and center panels,
the same interaction is used to generate both the folding model optical
potential and the inelastic transition. The curves follow the same scheme
as Figs.~7-9. Again, switching to high momentum matching reduces the 
density-dependent effects. Then the DBHF predictions appear to be the best,
and we will use this scheme in the rest of this paper. The right panels
show the Furnstahl and Wallace result. Here the dash-dot curve substitutes
the inelastic interaction in the calculation of the folding model
distortions. The good agreement with the DBHF (high momentum match) scheme
and the small difference between the two Furnstahl results would suggest
that using the correct inelastic interaction is more important than the 
M\"oller operator change to the distortions. 
Thus the DBHF scheme (together with the approximation illustrated by 
the short-dashed curve in Fig.~5) is a reasonable one. 
\vskip.4in
\noindent{\fta 7.~~Results of the DWIA Calculations}
\bigskip
After discussing the sensitivity of the predictions to some of the model ingredients, in
this section we will present and discuss the results from the DBHF model
interaction (with the larger effective masses shown in the right panel
of Fig.~6). We will begin with some additional comments about the calculation
of elastic scattering, then move to comments concerning the choice of the OBE
potential and some observations concerning a variety of natural parity
transitions in $^{16}$O and $^{40}$Ca at 200 MeV, and in $^{28}$Si at 180
MeV.

Figure 11 shows the folding model calculations for proton elastic
scattering from $^{16}$O and $^{28}$Si using the free (short dash), BHF
(long dash), and DBHF (solid) interactions. These curves were calculated
in the same manner as the curves in the center panels of Fig.~9 for
$^{40}$Ca. The local nuclear matter density was derived from the longitudinal
electron scattering form factor [50], as described in sect.~5. All 
of these measurements are at 200 MeV. The data for $^{28}$Si were taken 
in connection with an investigation of high spin states in that nucleus
[20]. The features here are similar to those described for $^{40}$Ca in 
the previous section. The DBHF calculations in their present form overestimate
the cross section and predict values for the analyzing power that are
generally too negative. In the analyzing power, we see a progression
from positive to negative values as the size of the density dependence
increases. This suggests that the total density dependence is too large,
an effect that may be associated with the infinite nuclear matter assumption
of our G-matrix calculation. 
\topinsert
\vbox to 3.6 truein {\vss
\hbox to 6.5 truein {\includegraphics{finfig11.ps}\hss}}
\noindent{\it Figure 11.}~~Measurements of the cross section and analyzing power
for $^{16}$O and $^{28}$Si at 200~MeV.
The folding model calculations are based on the 
free (short dash), BHF (long dash), and DBHF (solid) forms of the effective
interaction.            
\endinsert

It is clearly possible to achieve a better phenomenological description
of the elastic scattering measurements through the use of optical 
potentials with adjustable shapes. Such a calculation [65] for proton 
elastic scattering on $^{28}$Si at 180~MeV is shown by the dashed line
in Fig.~12. The parameters of this potential vary smoothly from 80~MeV
[65] up to 200~MeV [66]. Clearly the adjusted optical potential produces
a good fit to the data, but it is valid to ask whether the optimization
of the fit to the elastic data builds in any prejudice. As is the case
with most optical model work, measurements are missing at small angles
where the cross sections are largest.
Thus the adjustment weighs more heavily the larger angle data where the 
cross section has fallen a few orders of magnitude and coupling to the 
major reaction channels may already be important. 
\topinsert
\vbox to 3.6 truein {\vss
\hbox to 6.5 truein {\includegraphics{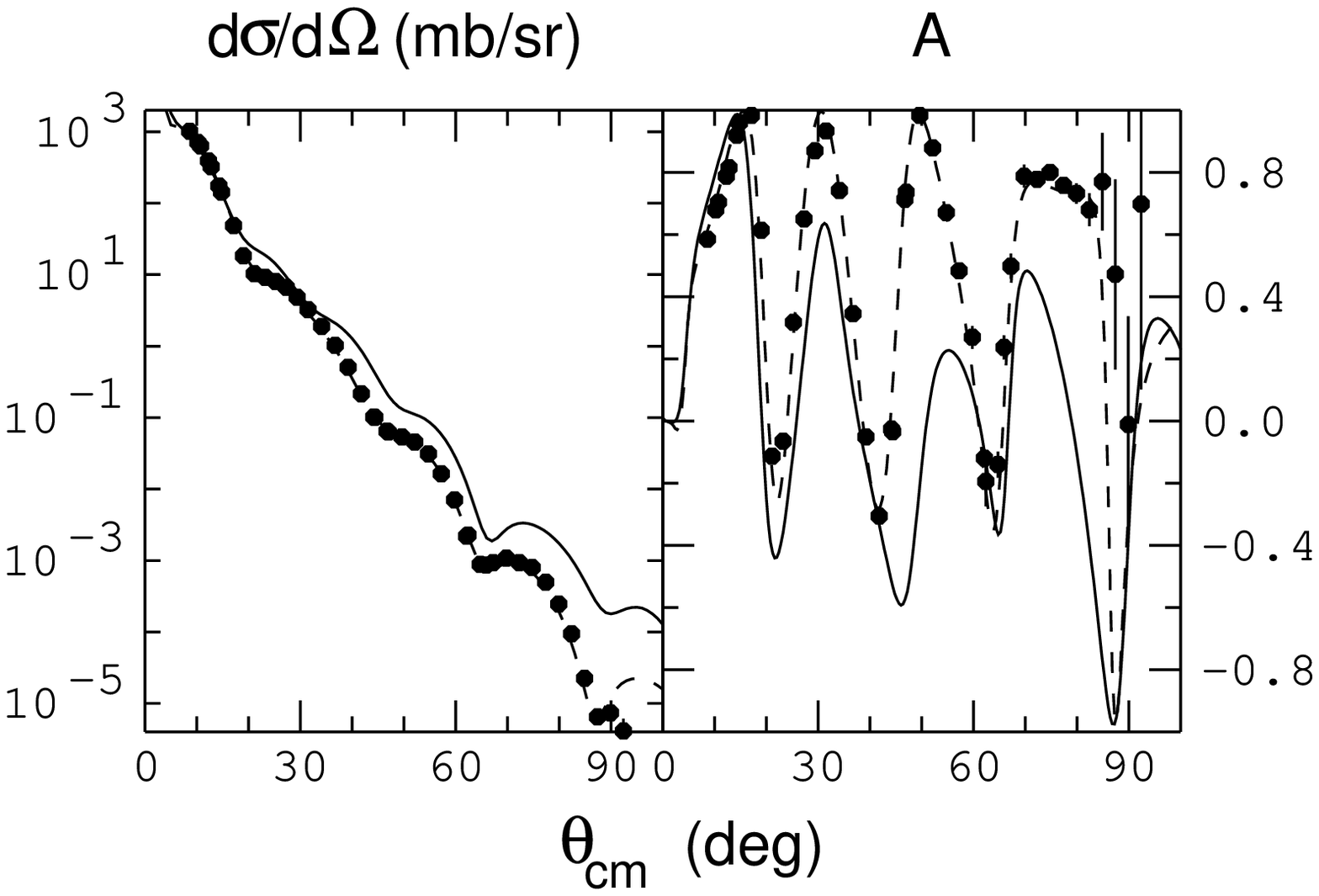}\hss}}
\noindent{\it Figure 12.}~~Measurements of the cross section and 
analyzing power for proton elastic scattering on $^{28}$Si at 180~MeV.
The solid curves show the results of a folding model calculation;
the long dashed curves are the final values from a phenomenological
optical potential fit to these data [57].
\endinsert
\topinsert
\vbox to 3.6 truein {\vss
\hbox to 6.5 truein {\includegraphics{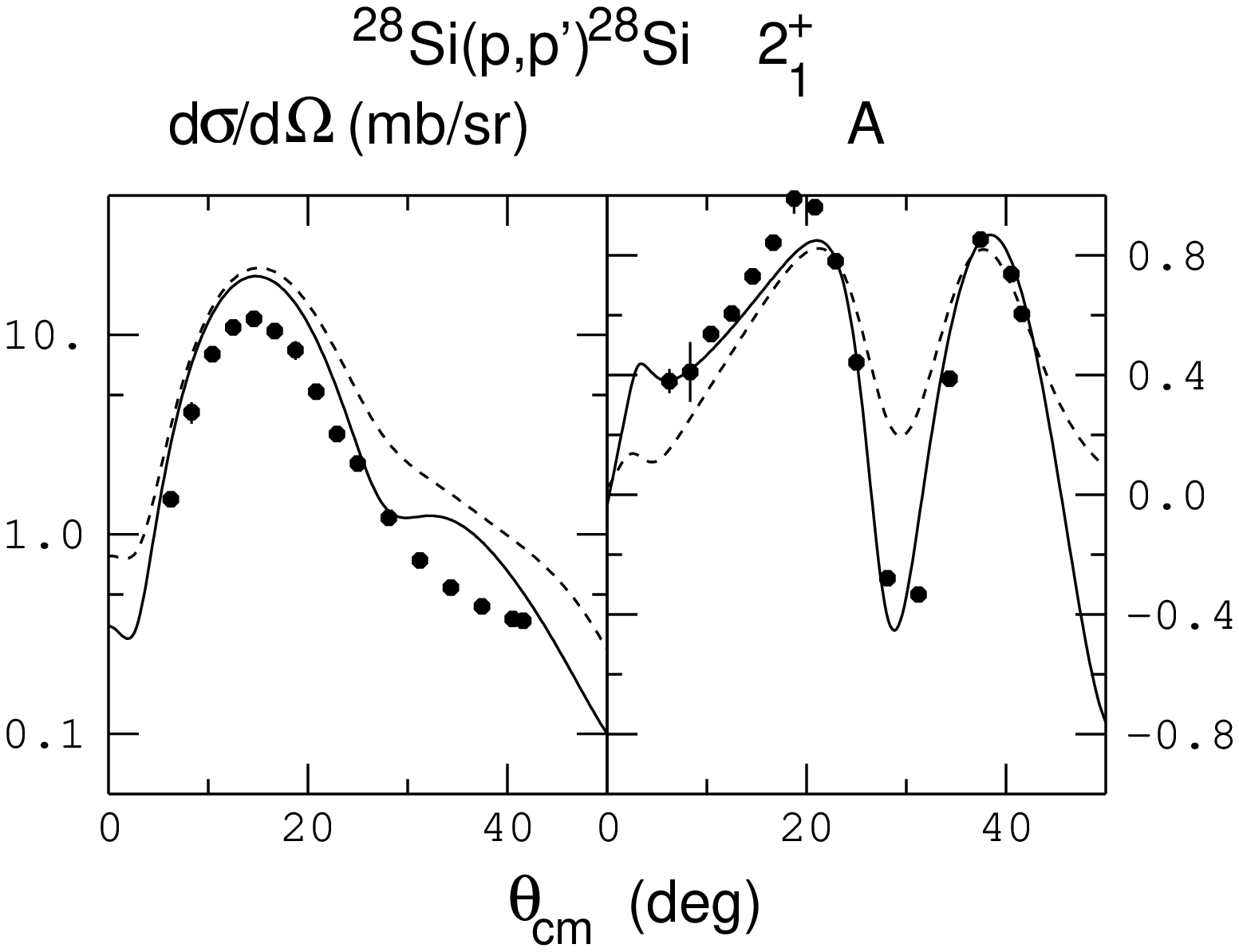}\hss}}
\noindent{\it Figure 13.}~~Measurements of the cross section and 
analyzing power for the $^{28}$Si(p,p$')^{28}$Si reaction to the
2$^+$ state at 1.779 ~MeV. The DBHF calculations are based on 
folding model (solid) and phenomenological best-fit optical potentials
(dashed).
\endinsert

One way to address this issue is to compare the two approaches to the 
distorted wave calculations for (p,p$'$) transitions. Fig.~13 shows
two DBHF calculations for the $^{28}$Si(p,p$'$)$^{28}$Si reaction to 
the 2$^{+}$ state at 1.779 MeV. The bombarding energy is 180 MeV and the 
form factor comes from the work of Chen [30]. The solid curves employ
folding model distortions, while the dashed curves are based on the 
best-fit phenomenological potential from Olmer [65]. The analyzing power
and cross section angular distributions both favor the use of folding 
model distortions. The best-fit optical potential gives cross section
predictions 
that fall too slowly at the larger angle. For the analyzing power,
both calculations do equally well in locating the diffraction pattern 
in angle, but the analyzing power predicted with the best-fit optical
potential is too positive. The same issues are also present for the other
strong (p,p$'$) transitions included in this study. Thus we are better
served by using distortions calculated within the folding model, despite
the concerns presented in the previous section.

Another point to realize is that with the measurement of a number of very
precise NN polarization observables near 200~MeV [67], the quality of the OBE
potential has improved. Because of the much larger theoretical
uncertainties introduced by the G-matrix calculations, it is not practical
to evaluate this issue by a direct comparison with (p,p$'$) data. Instead,
in Fig.~14 we show two calculations for the $^{40}$Ca(p,p$'$)$^{40}$Ca 
reaction to the 3$^{-}$ state at 3.736 MeV based on this work and the 
similar but earlier Bonn-B interaction [22].  For simplicity,
neither calculation contains density dependence.  
There are differences between the 
solid curve (Bonn-B) and the dashed curve (the OBE presented here),
particularly for the analyzing power. On the scale of our comparison to
(p,p$'$) data such differences matter, especially since they are systematic
across all transitions. Similar problems arise for other popular interactions,
such as the density-dependent interactions of von Geramb [24,25] and Nakayama
[26], as well as other free interactions such as the one from the Paris 
group [68]. As discussed previously [69], most modern interactions 
[70,34] are based
on the NN data base and phase shift analysis of the Nijmegen group [33]
and are correct to the extent that their selection of the NN data to be
considered is appropriate. In general, the older interactions are not as
faithful to the present NN scattering data, especially for measurements
involving polarization, and are likely to have larger discrepancies
when compared to (p,p$'$) measurements.
\topinsert
\vbox to 3.6 truein {\vss
\hbox to 6.5 truein {\includegraphics{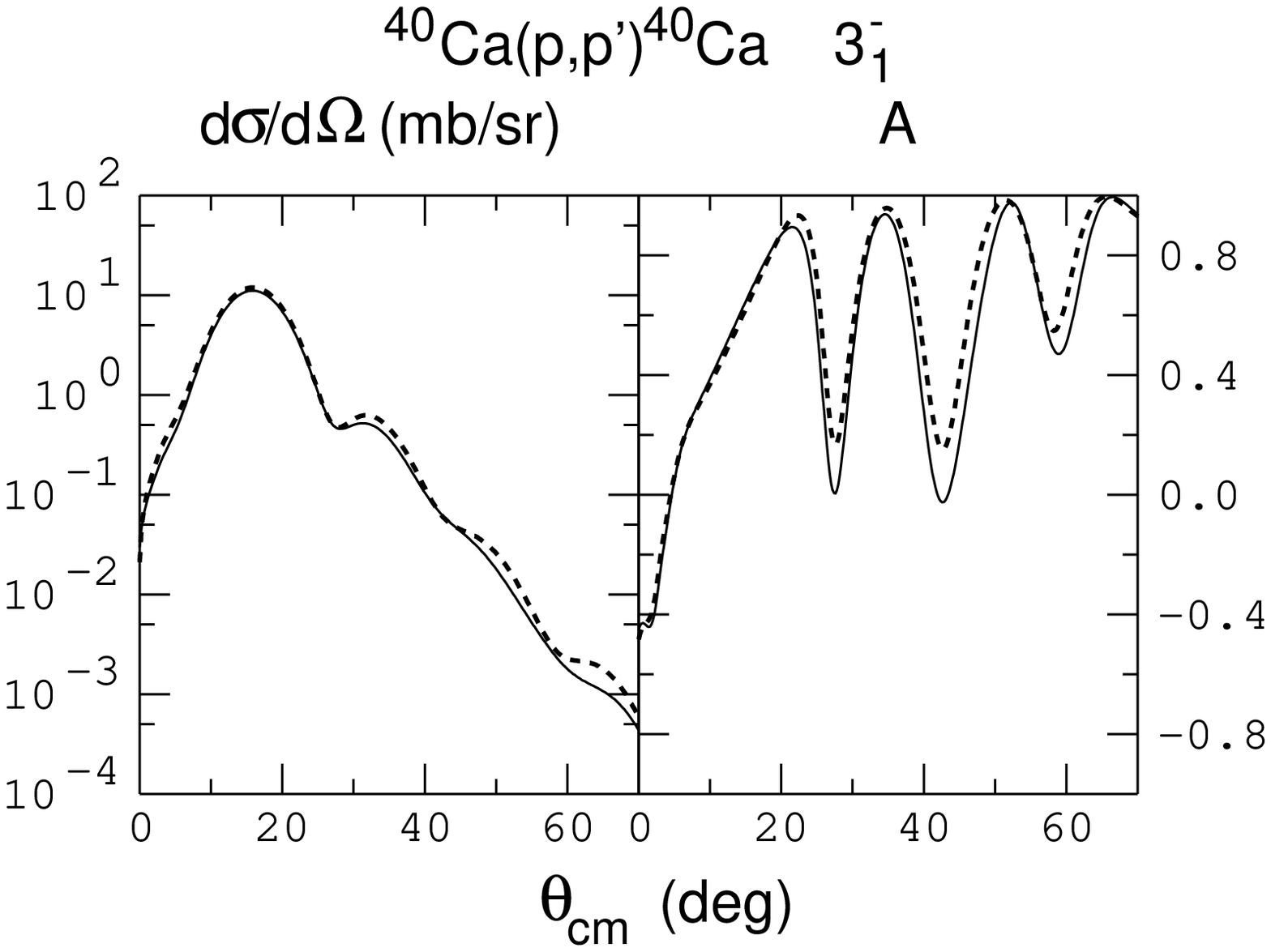}\hss}}
\noindent{\it Figure 14.}~~A comparison of two free interaction 
calculations for the $^{40}$Ca(p,p$')^{40}$Ca reaction to the 
3$^-$ state at 3.736~MeV. The solid line uses the Bonn-B interaction
and the dashed line uses the OBE potential of sect.~2.
\endinsert

Lastly, we will present calculations for a number of natural-parity
transitions in $^{16}$O, $^{28}$Si, and $^{40}$Ca in order to 
illustrate some additional properties and issues associated with our
model.                       

Figure 15 contains three strong collective states not previously shown
here,
\topinsert
\vbox to 3.6 truein {\vss
\hbox to 6.5 truein {\includegraphics{finfig15.ps}\hss}}
\noindent{\it Figure 15.}~~Measurements of the cross section and 
analyzing power for the (p,p$'$) reaction leading to the 2$^+$ and $3^-$
states in $^{16}$O at 6.917 and 6.130~MeV and the 2$^+$ state in 
$^{40}$Ca at 3.904~MeV.  The curves are based on the free (short
dash), BHF (long dash), and DBHF (solid) interactions.
\endinsert
along with calculations using the free (short dash), BHF (long dash), and
DBHF (solid) effective interactions. Presentation of all three interactions
again illustrates the effect of the major density-dependent contributions.  
In all cases the peak of the cross section is overpredicted by 
about 50\%.
A number of factors have been included that reduce the cross section, 
including relativistic kinematics for the transformation from t-matrix
to the Yukawa expansion, and the transformation of the interaction from
NN to N-nucleus coordinates described by Eq.~(14) of Love and Franey [21].
This has been an issue in other investigations, and has been handled in 
the empirical interaction of the Kelly group [29,30] by attenuating the free
interaction by factors of
about 0.8. This appears in Fig.~7 as a less negative
value of the Seifert central interaction at $k_F$=0~fm$^{-1}$ (the slope is
then adjusted to best match the same body of (p,p$'$) transitions that
we consider here). There the overestimate has been ascribed to an 
inadequacy in the local density approximation in which the density
dependence is scaled to the nuclear density at the coordinate of the 
projectile rather than considering some average over the locally changing
density that lies within the range of the NN interaction. 
At lower bombarding energies such overestimates of the cross section in
DWBA calculations have been attributed to distorted wave functions that are
too large in the nuclear interior because they ignore exchange non-localities 
[71]. Inclusion of the standard correction (scaled to the energy dependence
of the strenght of the central optical model potential) in fact increases
the cross section at the peak and does not, at least in this form, represent
a helpful way to handle this issue.

Generally, the shapes of the cross section angular distributions match
the data well. Of the three calculations shown, the best reproduction of the
analyzing power is clearly with the DBHF predictions. Thus a model that is
consistent with the bulk properties of nuclear matter is also best for
nucleon-induced reactions. 

For the weaker (p,p$'$) transitions, agreement with the calculations is a
bit more elusive. In Fig.~16, we again show the three 
types of calculations
for the 0$^+$ state in $^{28}$Si at 4.98~MeV and the 1$^-$ state in
$^{16}$O at 7.117~MeV, both non-collective transitions of low spin.
For the most part, the general features of the cross sections are 
reproduced, as would be expected since the transitions form factors are
taken from (e,e$'$) data.
\topinsert
\vbox to 3.6 truein {\vss
\hbox to 6.5 truein {\includegraphics{finfig16.ps}\hss}}
\noindent{\it Figure 16.}~~Measurements of the cross section and
analyzing power for the $0^+$ state in $^{28}$Si at 4.98~MeV and the
$1^-$ state in $^{16}$O at 7.117~MeV, along with calculations as
described in Fig.~15.
\endinsert

The higher spin states shown in Fig.~17 have simpler angular 
distributions. Here the quality of the reproduction is comparable
to the strong collective states shown in Fig.~15. All of the analyzing
powers show a falling angular distribution between 20$^\circ$ and 
40$^\circ$ that is well reproduced and shows little effect from the
density dependence. The overestimate of the cross section near the 
peak is still present.
\topinsert
\vbox to 3.6 truein {\vss
\hbox to 6.5 truein {\includegraphics{finfig17.ps}\hss}}
\noindent{\it Figure 17.}~~Measurements of the cross section
and analyzing power for the $4^+$ state in $^{16}$O at 10.356~MeV,
the $4^+$ state in $^{28}$Si at~4.62 MeV, and 
and the $5^-$ in $^{40}$Ca at~4.492 MeV, along with calculations as
described in Fig.~15.
\endinsert

None of the results so far discussed bear on the question of
whether the density dependence is correct, 
regardless of questions about the size of the interaction.
For this we need states whose transition densities peak
at nuclear densities that are
sufficiently different to make a 
meaningful comparison.  This opportunity is present
for $^{28}$Si, where the series of $2^+$ states shows the
needed variety.  In Fig.~18, the states considered in this
analysis (cleanly resolved in the experiment) are shown in
order of decreasing reaction radius.  This change is apparent
from the expansion of the diffraction pattern as the reaction
radius goes down.  If the reaction radius is larger, then one
would expect that most of the transition takes place in 
regions of the nucleus where the density is small and that
changes to the effective NN interaction would be modest.
This appears to be the case, as the size of the 
density-dependent modifications increases going from left
to right in Fig.~18.  This is true of both the cross section
and the analyzing power.  For the first state (4$^{\rm th}$
in excitation) all calculations
do a reasonable job capturing the features of the 
analyzing power angular distribution.                    
As the radius moves the transition to regions of higher nuclear 
density, the DBHF calculations continue to follow the measurements.
So the changing amount of density dependence needed to describe this
series of transitions is captured by our model.
\topinsert
\vbox to 3.6 truein {\vss
\hbox to 6.5 truein {\includegraphics{finfig18.ps}\hss}}
\noindent{\it Figure 18.}~~Measurements of the cross section
and analyzing power for three $2^+$ states in $^{28}$Si at 7.933,
1.779, and 8.259~MeV (in order of increasing nuclear density), 
along with calculations as described in Fig.~15.
\endinsert

In closing this section, we observe  
that a quantitative, relativistic two-body
force, combined with a Dirac approach to nuclear matter, 
provides a good description of
the main features of the (p,p$'$) data considered here.    
We also point out that 
a similar framework has proven successful
in microscopic nuclear structure calculations [72].
It must be stressed that 
 relativistic effects of moderate size, such as
those we obtain with the effective masses shown in the right panel of Fig.~6, are
crucial to obtain a satisfactory description of the main features of the 
data. 
\vskip.4in
\noindent{\fta 8.~~Conclusions}
\bigskip

The primary focus of this paper has been to establish a reliable
basis for inelastic scattering calculations with a medium-modified
effective NN interaction.  Thus we have concentrated on isoscalar,
natural-parity transitions where medium effects from the
established many-body models are known to be largest.  We have
reported our predictions for several such states obtained in 
(p,p$'$) scattering around 200~MeV.

Our starting interaction is based on a relativistic OBE model,
used together with the relativistic Thompson scattering equation.
The input OBE potential has been optimized to agree with the
best modern sets of phase shift solutions to the NN scattering
data base.  Medium effects are accounted for via a Dirac-Brueckner
many-body approach.  The proper handling of the nucleon in the
medium through a distorted Dirac wavefunction is well known to
provide a crucial saturation mechanism in nuclear matter.

We also report on the conversion of the G-matrix elements to a
Yukawa function representation of the effective NN interaction.
In this form, it is suitable for use in existing coordinate-space
DWIA calculations of (p,p$'$) or (p,n) reactions.  This
connection also makes a model-dependent separation of the effective
NN interaction into direct and exchange pieces.

For practical reasons, the distortions of the nucleon Dirac
wavefunction in nuclear matter are often handled in the so-called
effective mass approximation.  We present a detailed analysis of
the central issues involved in this approximation and show that,
for nucleon energies in the continuum, appropriate nucleon effective
masses are larger (namely, closer to the free space value) than those
obtained in typical nuclear matter calculations.

We compare our DBHF predictions with those from the relativistic model
of Furnstahl and Wallace and find similar quality for inelastic
scattering.  For elastic scattering, Furnstahl and Wallace achieve a
better agreement with data.  This is likely to come from the fact that
Furnstahl and Wallace treat nucleons as Dirac bound states in the
nucleus, rather than as infinite nuclear matter spinors (as we do).

Some systematic problems still remain with the cross section angular
distributions, where the size near the peak is overestimated by about
50\%.  This problem persists no matter which effective NN interaction
we use (except of course for phenomenological 
ones tuned to best fit the
cross section and analyzing power measurements), and may originate in
the choice of the DWIA as a reaction mechanism.

We report that a quantitative, relativistic two-body force, combined
with a Dirac approach (namely, under the conditions where nuclear
matter is well described), captures well the main features of the
(p,p$'$) data considered here.  We stress again that relativistic
effects of moderate size, such as those we obtain with the effective
masses shown on the right side of Fig.~6, are crucial to obtain a satisfactory
description of the main features of this data.  For future studies
involving the spin dependence of the in-medium NN interaction, this
model appears to be a satisfactory basis on which to proceed.
\bigskip
We acknowledge the assistance of Malcolm Macfarlane with the
transformation from G-matrix to effective NN interaction.  This
paper was produced with financial assistance from the US NSF under
grant NSF-PHY-9602872, 
a NATO travel grant 900235, and the University of Idaho.
\vskip.4in
\centerline{\fta Appendix}
\bigskip
This appendix details the transformation between an expansion of
the effective NN interaction using Yukawa functions for the direct
part and the t- or G-matrix elements generated by the in-medium
solution of the NN scattering potential problem.

The Yukawa functions in coordinate space follow the
work of Love and Franey [21],
and are $$t_C(r)=\sum_iV^C_iY(r/R_i),\quad\hbox{where }Y(x)=
e^{-x}/x\eqno{(A.1a)}$$ $$t_S(r)=\sum_iV^S_iY(r/R_i)\eqno{(A.1b)}$$
$$t_T(r)=\sum_iV^T_ir^2Y(r/R_i)\eqno{(A.1c)}$$ with the Fourier
transforms $$t_C(k)=4\pi\sum_i{V^C_iR^3_i\over 1+(kR_i)^2}
\eqno{(A.2a)}$$ $$t_S(k)=8\pi\sum_i{V^S_ikR^5_i\over (1+(kR_i)^2)
^2}\eqno{(A.2b)}$$ $$t_T(k)=32\pi\sum_i{V^T_ik^2R^7_i\over (1+
(kR_i)^2)^3}.\eqno{(A.2c)}$$

The S-matrix elements may be derived from the (complex) phase shifts
according to $$S_i=e^{i\delta_i}\eqno{(A.3a)}$$ if the phase is
uncoupled, or $$S_i=\left(\matrix{e^{2i\delta_1}\cos 2\epsilon&
ie^{i(\delta_1+\delta_2)}\sin 2\epsilon\cr ie^{i(\delta_1+\delta_2)}
\sin 2\epsilon& e^{2i\delta_2}\cos 2\epsilon\cr}\right)
\eqno{(A.3b)}$$ if
the phase shifts are coupled.  Here the quantum numbers $i$ include
the total angular momentum $J$, the total spin $S$, the total
isospin $T$, and the orbital angular momenta $\ell$ and $\ell '$.

The S-matrix elements are changed into T-matrix elements through
the relation $$S^{JST}_{\ell '\ell}(p)=\delta_{\ell '\ell}-i
\kappa\ t^{JST}_{\ell '\ell}(p).\eqno{(A.4)}$$  The coefficient
$\kappa =2\pi\mu p$ where $\mu$ is the reduced mass in the NN
system and $p$ is the NN center-of-mass momentum.

The closed form expression for the elements of the transform
matrix {\bf M} is $$t^{JST}_{\ell '\ell}={1\over 2p^2}\Bigl(
\delta_{\ell '\ell}\ g^{ST}_C(\ell ,p)\eqno{(A.5)}$$ $$-6
\delta_{\ell '\ell}\delta_{S1}\ (-)^{\ell +J}\left\{\matrix
{\ell& J& 1\cr1& 1& \ell\cr}\right\}\Bigl[ \sum_\lambda(2\lambda 
+1)\langle\lambda\ 1\ 0\ 0|\ell\ 0\rangle^2\left\{\matrix{\ell&
\ell& 1\cr 1& 1& \lambda\cr}\right\}\ g^T_S(\lambda ,p)\Bigr]$$
$$-(-)^{(\ell '-\ell)/2}2\sqrt{30}\delta_{S1}(-)^{\ell '+J+3}
\left\{\matrix{\ell '& J& 1\cr 1& 2& \ell\cr}\right\}\Bigl[
\sqrt{2\ell '+1}\langle\ell '\ 2\ 0\ 0|\ell\ 0\rangle{1\over p^2}
\bigl( g^T_T(\ell ,p)+g^T_T(\ell ',p)\bigr)$$ $$+\sqrt{30}
\sum_\lambda (2\lambda +1)\langle\lambda\ 1\ 0\ 0|\ell '\ 0
\rangle\langle\lambda\ 1\ 0\ 0|\ell\ 0\rangle\left\{\matrix{
\ell '& \ell& 2\cr1& 1& \lambda\cr}\right\}{1\over p^2}
g^T_T(\lambda ,p)\Bigr]\Bigr).$$  The sums on $\lambda$ run
from the maximum of $\ell -1$ or $\ell '-1$ to the minimum
of $\ell +1$ or $\ell '+1$.  The Yukawa coefficients are
contained in the expressions for the $g$ functions as $$g^{ST}_C
(\ell ,p)=\sum_\alpha V^C_\alpha R^{}_\alpha Q^{}_\alpha 
(y^{}_\alpha )\eqno{(A.6a)}$$ $$g^T_S(\ell ,p)=\sum_\alpha
V^S_\alpha R^{}_\alpha Q'_\alpha (y^{}_\alpha )\eqno{(A.6b)}$$
$$g^T_T(\ell ,p)=\sum_\alpha V^T_\alpha R^{}_\alpha Q''_\alpha
(y^{}_\alpha )\eqno{(A.6c)}$$ where the $Q_\alpha$ are
Legendre functions of the second kind and the derivatives are
with respect to $y_\alpha$ where $$y_\alpha =1+{1\over 2p^2_{}
R^2_\alpha}\eqno{(A.6d)}$$  The $V^C$, $V^S$ and $V^T$ are the
central, spin-orbit, and tensor Yukawa coefficients, and the
$R$ are their associated ranges.

The details of this transform were provided by Malcolm Macfarlane,
who derived them as part of an extended study of the off-shell
unitarity of the effective NN interaction [73].  While the version
here recognizes only one value of the momentum $p$, an extended
version exists that makes the connection to either a half-off-shell
or a fully-off-shell t- or G-matrix element.

\vskip.4in
\centerline{References}
\bigskip
\item{1.} K.A. Brueckner, C.A. Levinson, and H.M. Mahmoud,
Phys.\ Rev.\ {\bf 95}, 217 (1954).
\item{2.} H.A. Bethe, Phys.\ Rev.\ {\bf 103}, 1353 (1956).
\item{3.} J. Goldstone, Proc.\ R. Soc.\ (London) A {\bf 239}, 
267 (1957).
\item{4.} H.A. Bethe, Ann.\ Rev.\ Nucl.\ Sci.\ {\bf 21}, 93 
(1971).
\item{5.} M.I. Haftel and F. Tabakin, Nucl.\ Phys.\ {\bf 
A158}, 1 (1970). 
\item{6.} D.W.L. Sprung, Adv.\ Nucl.\ Phys.\ {\bf 5}, 225 
(1972).
\item{7.} M.R. Anastasio, L.S. Celenza, W.S. Pong, and C.M. 
Shakin, Phys.\ Rep.\ {\bf 100}, 327 (1983).
\item{8.} R. Brockmann and R. Machleidt, Phys.\ Lett.\ {\bf 
149B}, 283 (1984); Phys.\ Rev.\ C {\bf 42}, 1965 (1990).
\item{9.} C.J. Horowitz and B.D. Serot, Phys.\ Lett.\ {\bf 
137B}, 287 (1984); Nucl.\ Phys.\ {\bf A464}, 613 (1987).
\item{10.} B. ter Haar and R. Malfliet, Phys.\ Rep.\ {\bf 149}, 
207 (1987).
\item{11.} G.E. Brown, M. Buballa, Zi Bang Li, and J. Wambach,
Nucl.\ Phys.\ {\bf A593}, 295 (1995).
\item{12.} G.E. Brown and M. Rho, Phys.\ Rev.\ Lett.\ {\bf 66}, 2720 (1991);
Phys.\ Rep.\ {\bf 269}, 334 (1996).
\item{13.} T. Hatsuda and T. Kunihiro, Phys.\ Rep.\ {\bf 247}, 221 (1994),
and references therein.
\item{14.} E. Bleszynski, M. Bleszynski, and C.A. Whitten, Jr., Phys.\
Rev.\ C {\bf 26}, 2063 (1982).
\item{15.} J.M. Moss, Phys.\ Rev.\ C {\bf 26}, 727 (1982).
\item{16.} T.N. Taddeucci {\it et al.},\ Phys.\ Rev.\ Lett.\ {\bf 73},
3516 (1994).
\item{17.} W.G. Love, Amir Klein, M.A. Franey, and K. Nakayama, Can.\ 
J. of Phys.\ {\bf 65}, 536 (1987).
\item{18.} E.J. Stephenson and J.A. Tostevin, in ``Spin and Isospin in
Nuclear Reactions" eds. S.W. Wissink {\it et al.} (Plenum, 1991) p.~2.
\item{19.} H. Baghaei {\it et al.}, Phys.\ Rev.\ Lett.\ {\bf 69},
2054 (1992).
\item{20.} E.J. Stephenson, J. Liu, A.D. Bacher, S.M. Bowyer,
S. Chang, C. Olmer, S.P. Wells, and S.W. Wissink, \ Phys.\ Rev.\ 
Lett.\ {\bf 78}, 1636 (1997).
\item{21.} W.G. Love and M.A. Franey, Phys.\ Rev.\ C {\bf 24},
1073 (1981).
\item{22.} R. Machleidt, Adv.\ Nucl.\ Phys.\ {\bf 19}, 189
(1989).
\item{23.} James J. Kelly, {\it Program Manual for LEA}, 1995.
\item{24.} H.V. von Geramb, {\it The interaction Between Medium
Energy Nucleons in Nuclei - 1982} (AIP Conf.\ Proc.\ 97, 1983)
p.\ 44.
\item{25.} L. Rikus, K. Nakano, and H.V. von Geramb, Nucl.\ Phys.\ 
{\bf A414}, 413 (1984).
\item{26.} K. Nakayama and W.G. Love, Phys.\ Rev.\ C {\bf 38}, 51
(1988).
\item{27.} L. Ray, Phys.\ Rev.\ C {\bf 41}, 2816 (1990). 
\item{28.} R.J. Furnstahl and S.J. Wallace, Phys.\ Rev.\ C {\bf 47},
2812 (1993).
\item{29.} H. Seifert {\it et al.}\ Phys.\ Rev.\ C {\bf 47}, 1615
(1993).
\item{30.} Q. Chen, J.J. Kelly, P.P. Singh, M.C. Radhakrishna,
W.P. Jones, and H. Nann, Phys.\ Rev.\ C {\bf 41}, 2514 (1990).
\item{31.} Jian Liu, E.J. Stephenson, A.D. Bacher, S.M. Bowyer,
S. Chang, C. Olmer, S.P. Wells, S.W. Wissink, and J. Lisantti,
Phys.\ Rev.\ C {\bf 53}, 1711 (1996).
\item{32.} R. Machleidt, {\it One-boson exchange potentials and
nucleon-nucleon scattering}, Comp.\ Nucl.\ Phys.\ 2 -- Nucl.
Reactions, eds.\ K. Langanke, J.A. Maruhu, and S.E. Koonin
(Springer,New York, 1993) p.\ 1.
\item{33.} V.G.J. Stoks {\it et al.},\ Phys.\ Rev.\ C {\bf 49},
2950 (1994).
\item{34.} R. Machleidt, F. Sammarruca, and Y. Song, Phys.\ 
Rev.\ C {\bf 53}, 1483 (1996).
\item{35.} W. Legindgaard, Nucl.\ Phys.\ {\bf A297}, 429 (1978).
\item{36.} T. Cheon and E.F. Redish, Phys.\ Rev.\ C {\bf 39},
331 (1989). 
\item{37.} L.D. Miller and A.E.S. Green, Phys.\ Rev.\ C {\bf 
5}, 241 (1972).
\item{38.} J.D. Walecka, Ann.\ Phys.\ (N.Y.) {\bf 83}, 491 
(1974). 
\item{39.} L.G. Arnold, B.C. Clark, and R.L. Mercer, Phys.\ 
Rev.\ C {\bf 19}, 917 (1979).
\item{40.} R. Brockmann, Phys.\ Rev.\ C {\bf 18}, 1510 (1978).
\item{41.} C.J. Horowitz and B.D. Serot, Nucl.\ Phys.\ {\bf 
A368}, 503 (1981).
\item{42.} B.D. Serot and J.D. Walecka, Adv.\ Nucl.\ Phys.\ 
{\bf 16}, 1 (1986).
\item{43.} M. Jaminon, Nucl.\ Phys.\ {\bf A402}, 366 (1983). 
\item{44.} G.R. Satchler, {\it Direct Nuclear Reactions} 
(Oxford, 1983) p.\ 115.
\item{45.} M.A. Franey and W.G. Love, Phys.\ Rev.\ C {\bf 31},
488 (1985).
\item{46.} S. Karataglidis, P.J. Dortmans, K. Amos, and R. de
Swiniarski, Phys.\ Rev.\ C {\bf 52}, 861 (1995).
\item{47.} T. Cheon, K. Takayanagi, and K. Yazaki, Nucl.\ Phys.\ 
{\bf A437}, 301 (1985); {\bf A445}, 227 (1985).
\item{48.} T. Cheon and K. Takayanagi, Nucl.\ Phys.\ {\bf A455},
653 (1986).
\item{49.} A.K. Kerman, H. McManus, and R.M. Thaler, Ann.\ Phys.\ 
{\bf 8}, 551 (1959).
\item{50.} H. de Vries, C.W. de Jager, and C. de Vries, At.\ Data
and Nucl.\ Data Tables {\bf 36}, 495 (1987).
\item{51.} L. Ray and G.W. Hoffman, Phys.\ Rev.\ C {\bf 31}, 538 (1985).
\item{52.} S. Karataglidis, P.J. Dortmans, K. Amos, R. de
Swiniarski, Phys.\ Rev.\ C {\bf 53}, 838 (1996).
\item{53.} P.J. Dortmans, K. Amos, and S. Karataglidis, J. Phys.\ 
G {\bf 23}, 183 (1997).
\item{54.} M.A. Suhail, S.M. Saliem, and W. Haider, J. Phys.\ G 
{\bf 23}, 365 (1997).
\item{55.} S.P. Weppner, Ch.\ Elster, and D. H\"uber, Phys.\ Rev.\ 
C {\bf 57}, 1378 (1998).
\item{56.} E. Bauge, J.P. Delaroche, and M. Girod, Phys.\ Rev.\ C
{\bf 58}, 1118 (1998).
\item{57.} H.F. Arellano, F.A. Brieva, and W.G. Love, Phys.\ Rev.\ 
C {\bf 52}, 301 (1995).
\item{58.} R. Crespo, R.C. Johnson, and J.A. Tostevin, Phys.\ Rev.\ 
C {\bf 53}, 3022 (1996).
\item{59.} Ch.\ Elster, S.P. Weppner, and C.R. Chinn, Phys.\ Rev.\ 
C {\bf 56}, 2080 (1997).
\item{60.} H.O. Meyer {\it et al.}, Phys.\ Rev.\ C {\bf 23}, 616 (1981).
\item{61.} R.M. de Vries and N. Di Giacomo, J.\ Phys.\ G {\bf 7}, ?51 (1981).
\item{62.} H.O. Meyer and P. Schuandt, Phys.\ Lett.\ {\bf 107B}, 353 (1981).
\item{63.} J.A. Tjon and S.J. Wallace, Phys.\ Rev.\ C {\bf 35},
280 (1987); Phys.\ Rev. C {\bf 36}, 1085 (1987).
\item{64.} J.J. Kelly and S.J. Wallace, Phys.\ Rev.\ C {\bf 49},
1315 (1994).
\item{65.} C.A. Olmer {\it et al.},\ Phys.\ Rev.\ C {\bf 29}, 361
(1984).
\item{66.} J. Liu, Ph.D.\ thesis, Indiana University, 1995.
\item{67.} for example, F. Rathmann {\it et al.},\ Phys.\ Rev.\ C 
{\bf 58}, 658 (1998).
\item{68.} M. Lacombe {\it et al.}, Phys.\ Rev.\ C {\bf 21}, 861 (1980).
\item{69.} F. Sammarruca and E.J. Stephenson, Phys.\ Rev.\ C
{\bf 58}, 307 (1998).
\item{70.} R.B. Wiringa {\it et al.}, Phys.\ Rev.\ C {\bf 51}, 38 (1995).
\item{71.} F. Perey and B. Buck, Nucl.\ Phys.\ {\bf 32}, 353 (1962).
\item{72.} M.F. Jiang {\it et al.}, Phys.\ Rev.\ C {\bf 46}, 910 (1992).
\item{73.} M.H. Macfarlane and Edward F. Redish, Phys.\ Rev.\ C 
{\bf 37}, 2245 (1988).
\vfill
\end